\documentclass[]{aa} 
\usepackage{graphicx}
\PassOptionsToPackage{hyphens}{url}
\usepackage[varg]{txfonts}
\usepackage{multirow}
\graphicspath{{./Figures/}}
\usepackage{twoopt}
\usepackage{natbib}
\bibpunct{(}{)}{;}{a}{}{,} 
\usepackage[breaklinks=true]{hyperref} 
\makeatletter
\newcommandtwoopt{\citeads}[3][][]{\href{https://ui.adsabs.harvard.edu/\#abs/#3}%
{\def\hyper@linkstart##1##2{}%
\let\hyper@linkend\@empty\citealp[#1][#2]{#3}}}
\newcommandtwoopt{\citepads}[3][][]{\href{https://ui.adsabs.harvard.edu/\#abs/#3}%
{\def\hyper@linkstart##1##2{}%
\let\hyper@linkend\@empty\citep[#1][#2]{#3}}}
\newcommandtwoopt{\citetads}[3][][]{\href{https://ui.adsabs.harvard.edu/\#abs/#3}%
{\def\hyper@linkstart##1##2{}%
\let\hyper@linkend\@empty\citet[#1][#2]{#3}}}
\newcommandtwoopt{\citeyearads}[3][][]%
{\href{https://ui.adsabs.harvard.edu/\#abs/#3}
{\def\hyper@linkstart##1##2{}%
\let\hyper@linkend\@empty\citeyear[#1][#2]{#3}}}
\makeatother
\hypersetup{colorlinks = true, allcolors={cyan}} 
\newcommand{\W}{$\mathrm{W3(H_2O)}$}

\newcommand{\mc}{$\mathrm{CH_3CN}$}
\newcommand{\mck}[1]{$\mathrm{CH_3CN}\,(12_#1-11_#1)$}
\newcommand{\mckr}[2]{$\mathrm{CH_3CN}\,(12_K-11_K)\,K = #1-#2$}

\newcommand{\eg}    {e.g.,}

\newcommand{\jpb}   {Jy~beam$^{-1}$}

\newcommand{\kms}   {km~s$^{-1}$}

\newcommand{\mo}    {$M_{\sun}$}

\newcommand{\tq}{Toomre~$Q$}

\newcommand{\uv}{\textit{uv}}

\begin{document} 

   \title{Disk Kinematics and Stability in High-Mass Star Formation} 

   \subtitle{Linking Simulations \& Observations}

   \author{A.~Ahmadi\inst{1,2}
          \and
          R. Kuiper\inst{3,1}
          \and
          H. Beuther\inst{1}
          }
   \institute{Max Planck Institute for Astronomy, K\"onigstuhl 17,
              69117 Heidelberg, Germany, e-mail: \href{mailto:ahmadi@mpia.de}{\nolinkurl{ahmadi@mpia.de}}
  \and
Fellow of the International Max Planck Research School for Astronomy and Cosmic Physics at the University of Heidelberg (IMPRS-HD)
  \and
  Institute of Astronomy and Astrophysics, University of T\"ubingen, Auf der Morgenstelle 10, 72076, T\"ubingen, Germany
    }

  \authorrunning{A. Ahmadi et al.}
  \titlerunning{}
   
  \date{Accepted in September 2019}

  \abstract
   {In the disk-mediated accretion scenario for the formation of the most massive stars, high densities and accretion rates could induce gravitational instabilities in the disk, forcing it to fragment and produce companion objects. }
   {We investigate the effects of inclination and spatial resolution on observable kinematics and stability of disks in high-mass star formation.}
   {We study a high-resolution 3D radiation-hydrodynamic simulation that leads to the fragmentation of a massive disk. Using \emph{RADMC-3D} we produce 1.3~mm continuum and \mc\ line cubes at different inclinations. The model is set to different distances and synthetic observations are created for ALMA at $\sim$80~mas resolution and NOEMA at $\sim$0.3\arcsec.}
   {The synthetic ALMA observations resolve all fragments and their kinematics well. The synthetic NOEMA observations at 800~pc with linear resolution of $\sim$300~au are able to resolve the fragments, while at 2000~pc with linear resolution of $\sim$800~au only a single structure slightly elongated towards the brightest fragment is observed. The position-velocity (PV) plots show the differential rotation of material best in the edge-on views. A discontinuity is seen at a radius of $\sim$250~au, corresponding to the position of the centrifugal barrier. As the observations become less resolved, the inner high-velocity components of the disk become blended with the envelope and the PV plots resemble rigid-body-like rotation. Protostellar mass estimates from PV plots of poorly resolved observations are therefore overestimated. We fit the emission of \mck{K}\ lines and produce maps of gas temperature with values in the range of 100--300~K. Studying the Toomre stability of the disks, we find low $Q$ values below the critical value for stability against gravitational collapse at the positions of the fragments and in the arms connecting the fragments for the resolved observations. For the poorly resolved observations we find low $Q$ values in the outskirts of the disk. Therefore, despite not resolving any of the fragments, we are able to predict that the disk is unstable and fragmenting. This conclusion is true regardless of knowledge about the inclination of the disk.}
   {These synthetic observations reveal the potential and limitations to study disks in high-mass star formation with current (mm) interferometers. While the extremely high spatial resolution of ALMA reveals extraordinary details, rotational structures and instabilities within accretion disks can also be identified in poorly resolved observations.}
   
   \keywords{stars: formation -- 
                     stars: massive -- 
                     stars: kinematics and dynamics --
                     accretion, accretion disks --
                     methods: numerical --
                     techniques: interferometric
                     }
\maketitle
\section{Introduction}
High-mass stars ($M\gtrsim 8$~\mo) live short and violent lives, ejecting a significant amount of mechanical and radiative energy into the Universe, enriching the interstellar medium with the heavy material needed for the creation of organic life as we know it. They affect not only subsequent star and planet formation, but also the evolution of the galaxy. Despite their importance, the short-lived nature of their existence makes studying their birth and evolution extremely difficult. To make matters worse, high-mass stars are rare (\citeads{2001MNRAS.322..231K}; \citeads{2003PASP..115..763C}) and typically found at distances much further than low-mass star-forming regions. 

High-mass stars experience rapid Kelvin-Helmholtz contractions, allowing them to start burning hydrogen while still accreting material \citepads{1993ApJ...418..414P}. This lack of pre-main-sequence phase in the evolution of high-mass stars would mean that intense  (ionising) radiation from the protostar is present in the main accretion phase of the protostar. For many years, the effect of radiation pressure on dust was thought to hinder the formation of stars more massive than $\sim$40~\mo\ (\eg\ \citeads{1971A&A....13..190L};  \citeads{1974A&A....37..149K}; \citeads{1987ApJ...319..850W}). Such studies would adopt nearly dust-free cloud conditions and extremely high accretion rates to allow the build-up of enough material onto the protostar to even reach such stellar upper mass limits. Subsequent works showed that the unrealistic restrictions on the dust opacity and accretion rates can be removed if non-spherical accretion flows are considered (\eg\ \citeads{1989ApJ...345..464N}). In particular, magnetic fields, rotation or even simple contraction of the cloud could provide such deviations from spherical symmetry. In more recent years, two- and three-dimensional radiation-hydrodynamical simulations of collapsing cores have shown support for disk-mediated accretion in the formation of very massive stars, analogous to the formation of low-mass stars (\eg\ \citeads{2002ApJ...569..846Y}; \citeads{2009Sci...323..754K}; \citeads{2010ApJ...722.1556K}, \citeyearads{2011ApJ...732...20K}; \citeads{2013ApJ...772...61K}; \citeads{2016ApJ...823...28K}; \citeads{2016MNRAS.463.2553R}; \citeads{2018A&A...616A.101K}). In the disk-mediated accretion, the effect of radiation pressure would be reduced as the radiation could escape through the poles along the disk rotation axis, and the disk would be shielded from radiation due to its high optical depth.  

While the formation mechanism of high-mass stars through accretion disks is being comprehensively studied theoretically, the observational existence of disks around O- and B-type stars is still elusive. This is in part due to their scarcity, fast evolution, and the fact that high-mass stars form in dense distant clusters, making disentangling individual stellar contributions difficult (see review by \citeads{2018ARA&A..56...41M}). Perhaps the best observational evidence for the existence of disks is the ubiquitous observations of collimated bipolar outflows (\eg\ \citeads{2002A&A...387..931B}; \citeads{2009A&A...504..127F}; \citeads{2011A&A...530A..12L}; \citeads{2014prpl.conf..451F}; \citeads{2015MNRAS.453..645M}; \citeads{2019A&A...623A..77S}), which has also been predicted by theoretical models (\eg\ \citeads{2007prpl.conf..277P}; \citeads{2008ASPC..387..216B}; \citeads{2018A&A...620A.182K}). A review by \citetads{2016A&ARv..24....6B} summarises our current understanding of the properties of accretion disks around young intermediate- to high-mass stars. They refer to large structures for which a rotational velocity field could be established, extending thousands of au in space, as `toroids' which host a (proto)cluster of stars. Furthermore, with the advent of long-baseline interferometry, a hand-full of disk candidates around the most massive stars have recently been detected (\citeads{2015ApJ...813L..19J}; \citeads{2015MNRAS.447.1826Z}, \citeyearads{2019ApJ...872..176Z}; \citeads{2016MNRAS.462.4386I}, \citeyearads{2018ApJ...869L..24I}; \citeads{2016ApJ...823..125C}; \citeads{2017A&A...602A..59C}; \citeads{2018A&A...617A..89C}; \citeads{2018A&A...618A..46A}; \citeads{2018A&A...620A..31M}, \citeyearads{2019A&A...627L...6M}; \citeads{2019A&A...623A..77S}). 

Although disk-mediated accretion has solved the radiation-pressure problem, high accretion rates ($\gtrsim 10^{-4}$~\mo\,$\mathrm{yr}^{-1}$) through the disk are still required for the creation of the most massive stars. Gas densities needed to provide such high rates of accretion could induce gravitational instabilities in the disk forcing it to fragment and produce companion objects \citepads{2006MNRAS.373.1563K}. While high densities can induce instabilities in the disk, thermal gas pressure and the shear force as a result of differential rotation of material in the disk can provide added stability against local collapse. The balance between these forces ultimately determines the fate of the disk. Originally introduced by \citetads{1960AnAp...23..979S} and later quantified by \citetads{1964ApJ...139.1217T}, a disk in Keplerian rotation is unstable against axisymmetric gravitational instabilities when the \tq\ parameter
\begin{equation} \label{e: Toomre}
  Q\equiv\frac{c_s\,\Omega}{\pi\,G\,\Sigma} < 1,
\end{equation}
where the stabilising effect of pressure is accounted for in the equation of sound speed $c_s$, shear is considered in the epicyclic frequency (or angular velocity) $\Omega$ of the disk, and the self-gravitational force is represented as the surface density $\Sigma$ of the disk.

Disks are prone to fragmentation when their mass exceeds  50\% of the mass of the entire system \citepads{2010ApJ...708.1585K}. Numerical simulations have found that disk fragmentation can result in the creation of short-period binaries on the scales of hundreds of au (\citeads{2017MNRAS.464L..90M}; \citeyearads{2018MNRAS.473.3615M}). The highest-resolution observations with the Atacama Large Millimeter Array (ALMA) that are able to resolve such structures on few hundred au scales are also starting to see fragmentation of the disks on these scales (\citeads{2018ApJ...869L..24I}). Considering the large fraction ($>70\%$) of OB stars that are found in close binary systems through radial velocity surveys \citepads{2012MNRAS.424.1925C}, it is important to understand the role of disk fragmentation in affecting the observed stellar mass distribution.

As we obtain the observational capabilities to resolve high-mass protostellar disks, it is important to create frameworks with which to study their properties. Synthetic observations play a crucial role in this effort as they not only allow for meaningful predictions to be made from theoretical models, but also provide valid interpretations for observations (for a review, see \citeads{2018NewAR..82....1H}). To date, only a few works have created synthetic observations from radiation-hydrodynamic simulations of high-mass star formation with the aim of providing predictions for long-baseline interferometry (\citeads{2007ApJ...665..478K}; \citeads{2017MNRAS.471.4111H}; \citeads{2018MNRAS.473.3615M}). Due to the computational cost of running numerical simulations and radiative transfer models for a wide range of parameters, such studies have not been very comprehensive. \citetads{2019MNRAS.482.4673J} were able to explore a larger parameter space by adopting a semi-analytical treatment of the density, temperature, and velocity fields with prescriptions for molecular freeze-out and thermal dissociation. Their synthetic dust continuum and molecular line observations probe different disk masses, distances, inclinations, thermal structures, dust distributions, and number and orientation of
spirals and fragments. 

In the following work, we aim to study the kinematics and stability of high-mass protostellar disks in detail by creating synthetic observations for the highest resolution 3D radiation-hydrodynamic simulations that lead to the fragmentation of a massive disk. In particular, we study the effect of inclination and spatial resolution on the derived disk and protostellar properties. To investigate the effects of spatial resolution, we assume a close distance of 800~pc and a more typical distance of 2~kpc, and create synthetic observations in the millimeter regime for ALMA at long baselines reaching very high resolutions ($\sim$80~mas) and for the NOrthern Extended Millimeter Array (NOEMA) with 6 antennas covering much shorter baselines currently providing the highest resolutions possible ($\sim$0.3\arcsec) in the northern sky. In this way, we can study the properties of high-mass protostellar disks when they are resolved as well as unresolved disks which resemble toroidal-like structures often seen in observations of intermediate- to high-mass stars. The motivation for this work is in part because we have observed the largest sample of 20 high-mass young stellar objects in the northern sky with NOEMA \citepads{2018A&A...617A.100B}\footnote{\url{http://www.mpia.de/core}} and aim to study the stability of the candidate disks and characterise their properties in a follow-up paper. It is therefore paramount to understand the observational limitations in detail and benchmark our methods. From the analysis of synthetic observations, not only do we learn about the observational limitations but will also be able to use the observations to their full potential and gain higher confidence in the conclusions. 

The paper is organised as follows. Sect.~\ref{s: sims} outlines the details of the simulations. The radiative transfer modelling scheme is summarised in Sect.~\ref{s: radiative_transfer}. In Sect.~\ref{s: synth_obs} we describe how the synthetic observations for ALMA and NOEMA were created. In Sect.~\ref{s: analysis_results} we summarise our analysis and results, divided into subsections describing both the continuum (Sect.~\ref{ss: continuum}) and gas kinematics (Sect.~\ref{ss: kinematics}), temperature distribution (Sect.~\ref{ss: temperature}), mass estimates from dust and position-velocity diagrams (Sect.~\ref{ss: masses}), and the Toomre stability of the disks (Sect.~\ref{ss: toomre}). A summary of the work and main conclusions can be found in Sect.~\ref{s: conclusions}.

\section{Simulations} \label{s: sims}
The numerical simulation starts from a collapsing reservoir of dusty gas and follows the evolution of the system including the formation and evolution of a central massive protostar and the formation and fragmentation of its surrounding accretion disk. A theoretical investigation of this simulation (plus a convergence study) will be presented in Oliva et al.~(in prep.) and covers an analysis of the disk's stability against non-axially symmetric features such as spiral arms, an analysis of its fragmentation probability, a determination of the fragment statistics, their masses, and orbital parameters. For the purpose of the synthetic observations presented herein, we summarise in the following the covered physics in the simulation, the limitations of the model, the physical initial conditions, and the numerical grid setup, highlighting the uniqueness of the extremely high spatial resolution of the disk fragmentation simulation. We refer to \citetads{2011ApJ...732...20K}, \citetads{2017MNRAS.464L..90M}, and \citetads{2018MNRAS.473.3615M} for previous three-dimensional studies of disk evolution around high-mass protostars that use the same numerical framework.

For the hydrodynamic evolution, we utilise version 4.1. of the open source code \emph{Pluto} (\citeads{2007ApJS..170..228M}; \citeads{2012ApJS..198....7M}). Self-gravity of the gas is included via the \emph{Haumea} module (\citeads{2010ApJ...722.1556K}; \citeads{2011ApJ...732...20K}). Stellar irradiation and disk cooling are modelled using the hybrid radiation transport module \emph{Makemake} (\citeads{2010A&A...511A..81K}; \citeads{2013A&A...555A...7K}), updated to a so-called two-temperature scheme \citepads{2011A&A...529A..35C}. The first application of this scheme is presented in \citetads{2018A&A...616A.101K} and \citetads{2018A&A...618A..95B}. The evolution of the central protostar follows the tracks by \citetads{2009ApJ...691..823H}. Details on the numerical implementation and coupling to the hydrodynamics are given in \citetads{2010ApJ...722.1556K} and \citetads{2013ApJ...772...61K}.

The simulation focusses on the formation, evolution, and fragmentation of the circumstellar accretion disk. In order to be able to achieve the required high spatial resolution (discussed in detail below), we have to neglect other potential feedback effects from the forming protostar. First, radiation forces are not taken into account. Due to the fact that we analyse a snapshot of the simulation, in which the central protostar has reached 10~\mo, radiation forces are indeed much lower than gravity and centrifugal forces. Second, photoionisation feedback and the formation of an HII region is not taken into account. Again, those physics take place at a later evolutionary stage than considered here, namely after the contraction of the central protostar to the zero age main sequence, when it starts to emit copious radiation in the EUV regime. With respect to feedback of radiation and pressure and photoionisation, we refer to \citet{2018A&A...616A.101K}. Third, no feedback by protostellar outflows is taken into account in this simulation unlike in our previous feedback studies (\citeads{2015ApJ...800...86K}; \citeads{2016ApJ...832...40K}). The injection of a fast outflow would reduce the numerical timestep per main iteration due to the CFL condition \citepads{1967IBMJ...11..215C} of the explicitly solved hydrodynamics. This would make the simulation computationally too expensive and modelling of the entire disk fragmentation phase would become infeasible. From a physical point of view, the outflow itself will only marginally impact the disk physics by lowering the optical depth in the bipolar outflow cavity. The most severe limitation of the current disk fragmentation model is the negligence of large-scale magnetic fields pervading the initial collapsing mass reservoir. Although we have previously used the same numerical framework to model the collapse of magnetised high-mass star forming regions \citepads{2018A&A...620A.182K}, the effect of magnetic fields are neglected here. The collapse of a magnetised mass reservoir would not only lead to the launching of fast, collimated jets and wide-angle outflows, but actually also impact the morphology and dynamics of especially the innermost disk structure via additional magnetic pressure and angular momentum transport. The final impact of magnetic fields on disk fragmentation has to be investigated in future studies.

The initial mass reservoir of the simulation is described by a spherically symmetric mass density profile $\rho(r) \propto r^{-1.5}$ and an axially symmetric rotation profile $\Omega(R) \propto R^{-0.75}$, where $r$ and $R$ denote the spherical and cylindrical radii, respectively. The outer radius of the reservoir is set to $0.1 \mbox{ pc}$ and its total mass to $200 \mbox{ M}_\odot$. The corresponding free-fall time is $t_\mathrm{ff} \approx 37 \mbox{ kyr}$. The ratio of rotational energy to gravitational energy of the initial mass reservoir is set to 5\%, corresponding to an angular rotation frequency normalised to a value of roughly $10^{-10} \mbox{ Hz}$ at $R = 10 \mbox{ au}$. The temperature of the gas and dust is initially set to $10 \mbox{ K}$.

\begin{figure}
   \centering
   \includegraphics[width=\hsize]{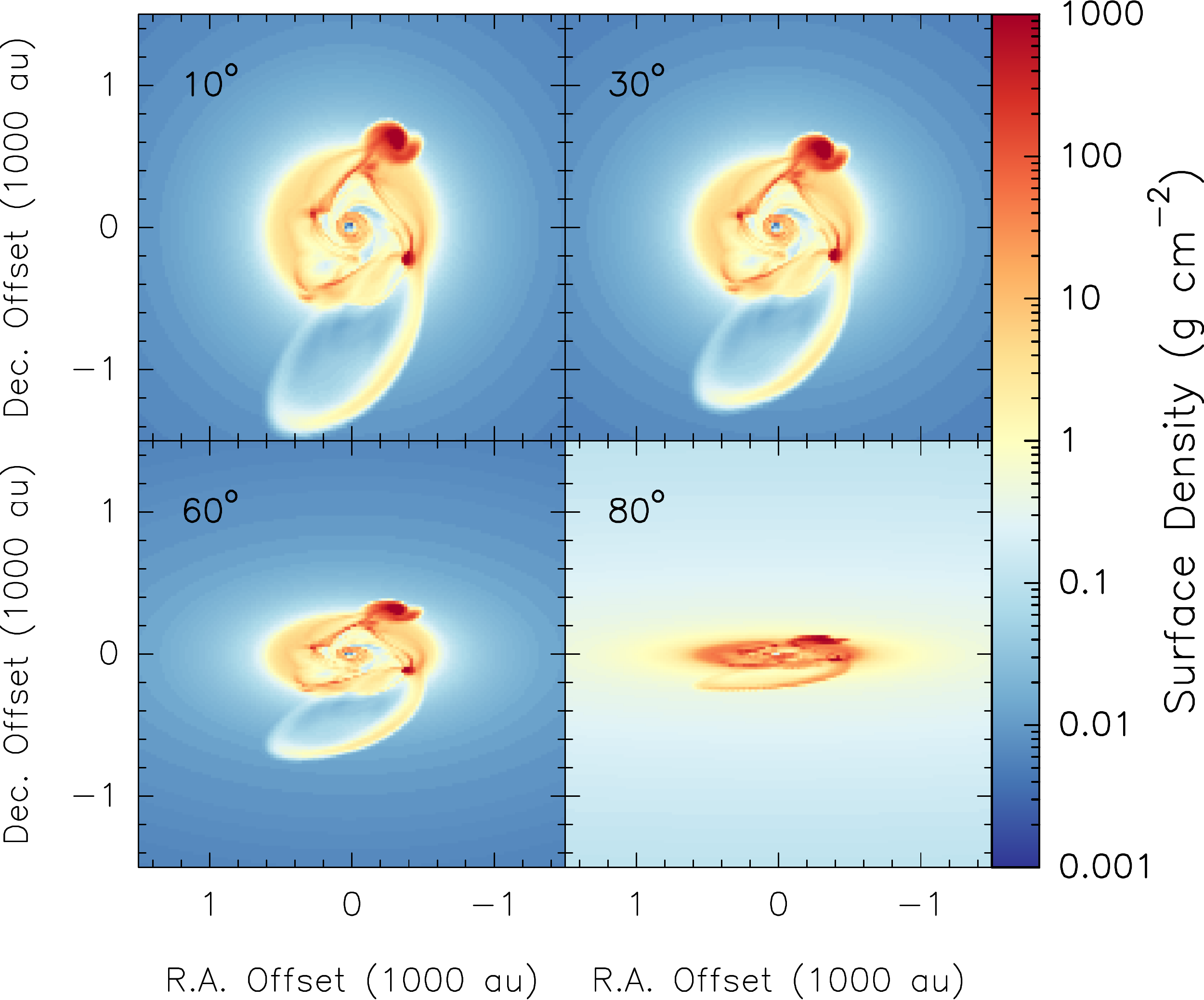}
      \caption{Disk surface density extracted from the numerical simulation at a snapshot of 12~kyr, inclined to 10\degr, 30\degr, 60\degr, and 80\degr\ as denoted in the panels.}
         \label{f: sim_sigma}
   \end{figure}

The uniqueness of this simulation is based on its extremely high spatial resolution of the forming accretion disk and its fragments. The gravito-radiation-hydrodynamical evolution of the system is computed on a numerical grid in spherical coordinates. In order to achieve a high dynamic range, we refine the grid cells in the radial direction toward the inner disk rim and in the polar direction toward the disk midplane. The refinement is done such that the grid cells in the disk midplane (at $\theta = 90\degr$) have the same extent in the radial, the polar, and the azimuthal direction ($\Delta r = r \Delta \theta = r \Delta \phi$). The radial extent of the inner sink cell is $30 \mbox{ au}$. The outer radius of the computational domain is $0.1 \mbox{ pc}$. In the polar direction, the grid extends from the upper to the lower polar axis ($0\degr \le \theta \le 180\degr$). In the azimuthal direction, the grid covers the full angle range as well ($0\degr \le \phi \le 360\degr$). The number of grid cells in the radial, the polar, and the azimuthal directions are $268 \times 81 \times 256$, respectively. This grid configuration results in a spatial resolution of the disk's midplane layer as function of distance to the host star $r$ of
\begin{equation}
\Delta x (r) \approx 0.025 \mbox{ au} \times r / (1 \mbox{ au}).
\end{equation}
Hence, at the inner rim of the computational domain at $30 \mbox{ au}$, the disk is resolved down to $0.75 \mbox{ au}$. We analyse the disk structure at 12~kyr after the onset of collapse when the disk has grown to $\approx 1000 \mbox{ au}$ (see Fig.~\ref{f: sim_sigma}). At this outer disk edge, the disk is resolved down to $25 \mbox{ au}$. In conclusion, the circumstellar disk (defined as gas with densities higher than $10^{-16} \mbox{ g cm}^{-3}$) is represented by more than a million grid cells in total. Both important physical length scales associated with disk evolution, namely the disk's pressure scale height $H$ (which is related to disk cooling) and the disk's Jeans length $\lambda_\mathrm{J}$ (which is related to the fragmentation process of the disk) are resolved. In comparison, to achieve the same dynamic range on a Cartesian grid with adaptive-mesh-refinement or nested levels, 15 levels of refinement would be required. This is calculated as follows: to resolve the entire computational domain with $L_\mathrm{box}=2\times0.1$~pc down to a minimum spatial resolution of $\mathrm{dx}=0.75$~au requires $2^n=L_\mathrm{box}/\mathrm{dx}$, so that the levels of refinement $n\sim15.7$. 

Besides the high spatial resolution, another important and unique feature of this simulation is that no subgrid models such as sink particles are used to represent fragments in the disk. A strongly self-gravitating subregion of the disk can undergo local collapse, and fragmentation barriers such as the required disk cooling and disk shear timescale are self-consistently taken into account due to the fact that the model solves for the hydrodynamics, radiation transport, and self-gravity, while resolving the pressure scale height and Jeans length.The high spatial resolution of the grid allows us to even resolve the hydrostatic physics of the forming first Larson cores, at least in the inner parts of the disk where fragmentation occurs first; fragments, which are expelled from the disk toward larger radii ($r > 1000 \mbox{ au}$) are not resolved first Larson cores anymore. Those first Larson cores, where self-gravity is in balance with internal thermal pressure, further interact with the large-scale accretion disk in which they are embedded. They accrete gas and dust from their own smaller-scale accretion disks (discussed later in Sect.~\ref{ss: kinematics}). The fragments migrate due to angular momentum exchange with the disk via gravitational torques, and sometimes merge, losing parts of their outer atmospheres back into the disk phase. Some fragments even get fully disrupted in the case of, for example, interactions with a spiral arm.

From this simulation, we take a snapshot at 12~kyr after the onset of collapse, when the host star has grown to 10~\mo. While at this snapshot the protostar would be classified as a B-type star, the setup of the simulation is valid for the formation of a more massive O-type star as it will continue to accrete through the disk and the migration of the fragments which induces episodic accretion onto the central protostar. Therefore, we expect this snapshot to also be representative of a slightly later evolutionary phase in which the protostar will be classified as an O-type star. In the simulations of  \citetads{2018MNRAS.473.3615M} which follow a similar numerical setup as ours and reach stellar masses as high as 35~\mo, the effects of stellar irradiation and the initial angular velocity distribution are tested. In all cases, the disk forms spiral arms that fragment due to gravitational instabilities as in the simulation analysed in this work. What the change in the initial angular velocity distribution affects is the disk size and the temporal evolution of the disk to star mass ratio, but the fact that the disk would eventually fragment remains unchanged. Similarly, the adoption of a steeper density distribution as done in some other works (e.g., $\rho(r) \propto r^{-2}$ in \citeads{2013ApJ...772...61K} and \citeads{2018A&A...616A.101K}) would make the disk initially more stable as more mass would be concentrated in the center rather than distributed in the envelope/disk and this lower disk-to-star mass ratio would make the disk more Toomre-stable initially, but not enough to hinder spiral arm formation and ultimately the fragmentation of the disk \citepads{2011ApJ...732...20K}. Therefore, the initial conditions we have chosen for this simulation are reasonable for the formation of massive protostar(s) and the study of disk fragmentation. 

Figure~\ref{f: sim_sigma} shows the surface density map extracted from the numerical simulations and inclined to $10\degr$, $30\degr$, $60\degr$, and $80\degr$, where $0\degr$ denotes a face-on view and $90\degr$ denotes an edge-on view of the circumstellar disk. To extract the surface density map, we have made use of the pressure scale height
\begin{equation}
 H(R)\approx \frac{R \,c_s}{v_\mathrm{KSG}},
\end{equation}
where $c_s$ is the sound speed and $v_\mathrm{KSG}$ the `Keplerian' velocity taking into account the central host mass as well as the disk mass interior to that radius $R$. The surface density is then calculated via 
\begin{equation}
  \Sigma(x, y) \approx \rho(x, y) \, H(R),
\end{equation}
where $\rho(x, y)$ is the gas mass density of the midplane. Since the calculation of the pressure scale height is only meaningful inside the disk region, the high column densities seen in the outskirts of the maps which are especially visible in the 80$\degr$ inclined map are not valid. The fragmented accretion disk of the snapshot at 12~kyr is further analysed by radiative transfer post-processing, as described in the following section.

\section{Radiative Transfer Post-Processing} \label{s: radiative_transfer}

The simulation snapshot under investigation consists of a central protostar with 10~\mo, a radius of $\approx 9 \mbox{ R}_\odot$, and an effective surface temperature of $17\,700 \mbox{ K}$, surrounded by a fragmented accretion disk of $\approx 2000 \mbox{ au}$ in diameter, harbouring spiral arms and a number of first Larson cores. We used the open source software \emph{RADMC-3D} \citepads{2012ascl.soft02015D} to create synthetic dust continuum observations at 1.3~mm and molecular line emission for the $12_K-11_K, K = 0-8$ rotational transitions of \mc. We will present the generated models in subsequent sections and in the following paragraph summarise the parameters used in \emph{RADMC-3D}. 

Gas and dust densities, temperature, and gas velocity are extracted from the hydrodynamics simulation snapshot. Gas and dust temperature are assumed to be in equilibrium, as it is done for the radiation-hydrodynamics simulation. Dust opacities are taken from \citetads{1993ApJ...402..441L}, used also for the continuum radiation transport in the hydrodynamics simulation. The \mc\ abundance (with respect to H$_2$) is set to $10^{-10}$ for temperatures $\le$90~K, $5\times10^{-9}$ for temperatures $>$90~K, and $10^{-8}$ for temperatures $>$100~K based on \citetads{2004MNRAS.354.1133C} and \citetads{2014A&A...563A..97G}. The spectral width of the ray-tracer is set to $\pm 247.8$~\kms\ around the central frequency of the non-Doppler-shifted line profile. The spectral resolution is set to 1400 frequency bins. The spatial resolution is $400 \times 400$ pixels for a region of the inner $8400 \mbox{ au} \times 8400 \mbox{ au}$ around the protostar. The inclination-dependent output of this radiative transfer post-processing task, mimicking an observation by a perfect telescope, is further processed to generate synthetic observational data cubes. Details are given in the following section.

\begin{table*}[!ht]
\renewcommand{\arraystretch}{1.3}
\caption{Details of synthetic observations.}
\centering
\label{t: config_specs}
\begin{tabular}{lcccccc}
\hline\hline
Interferometer & Distance  & Synthesised beam &  Linear resolution   & Continuum rms\tablefootmark{a} & Line rms\tablefootmark{b} \\
 & (pc) & (\arcsec $\times$\arcsec, PA) &  (au) & (m\jpb)&   (m\jpb)  \\
\hline
\multirow{ 2}{*}{ALMA} & 800 & 0.09\arcsec $\times$ 0.06\arcsec, 88\degr & 60 & 0.25 & 0.17  \\
 & 2000 &  0.07\arcsec $\times$ 0.06\arcsec, 90\degr & 130 & 0.06 & 0.04 \\
\hline
\multirow{ 2}{*}{NOEMA} & 800 & 0.44\arcsec $\times$ 0.34\arcsec, 44\degr & 312 & 0.60 & 1.08 \\
 & 2000 & 0.44\arcsec $\times$ 0.34\arcsec, 44\degr & 780 & 0.54 & 0.78 \\
\hline
\end{tabular}
\tablefoot{Values are averages for 10\degr, 30\degr, 60\degr, and 80\degr\ inclinations. \\
\tablefoottext{a}{The rms noise in the emission-free region.} \\
\tablefoottext  {b}{The rms noise in the emission-free region in the channel that has the peak of emission for \mck{4}, therefore the unit is per 0.35 \kms\ channel.}}
\end{table*}

\section{Synthetic Observations} \label{s: synth_obs}
We shift the inclined post-processed snapshot of the model simulations at 12~kyr in both dust continuum and lines to two different distances of 800 and 2000~pc by dividing the fluxes by a factor of distance squared and adjusting the pixel resolution accordingly. The generation of the synthetic observations for ALMA and NOEMA are described in detail in the following subsections. 

\subsection{ALMA}
To create synthetic ALMA observations, we use the ALMA simulator task \textit{simalma} within the Common Astronomy Software Applications (\emph{CASA}) software package\footnote{\url{http://casa.nrao.edu/}} to create visibility files. The model simulations adjusted to the desired distance and in units of Jansky per pixel are used as the `sky model'. The location of the sky model is set to a generic position in the southern sky, with the coordinates of the center of the map at $\alpha$(J2000) =  24$^{\rm h}$ 00$^{\rm m}$ 00$^{\rm s}$, $\delta$(J2000) = -35$^{\rm d}$ 00$^{\rm h}$ 00$^{\rm s}$. The observed sky frequency is set to 220.679~GHz with the widths of the channels for the line cubes set to 260.3~MHz ($\sim$0.35~\kms). We use a combination of two ALMA Cycle 5 configurations, 5-6 and 5-9, to reach a desired angular resolution of $\sim$0.08\arcsec. The configuration files containing the information about the distribution of the antennas were obtained within \emph{CASA} (version 5.1.2-4) and were labeled `alma.cycle5.6.cfg' and `alma.cycle5.9.cfg' for configurations 5-6 and 5-9, respectively. In Cycle 5, ALMA made use of 43 antennas in the 12-m arrays. The \uv-coverage of the ALMA array in configuration 5-6 has baselines in the range of 15~m to 2.5~km, and in configuration 5-9 covers a range of 375~m to 13.9~km. We set the integration time of the synthetic observations to 30-second intervals for a total time of 18 minutes in configuration 5-6 and 35 minutes in configuration 5-9. Figure~\ref{f: uv_coverage} shows the \uv-coverage of the synthetic observations. Thermal noise was automatically added to the observations assuming precipitable water vapour of 0.5~mm. 

The images were generated in \emph{CASA} using the \textit{clean} task with the \textit{h\"ogbom} algorithm \citepads{1974A&AS...15..417H} and a natural weighting. Considering the large number of channels in the data cubes (1400), we \textit{clean} the line cubes using a circular mask spanning the extent of the input sky model to make the imaging process run faster. The images at 800 and 2000~pc have synthesised beam sizes of 0.09\arcsec $\times$ 0.06\arcsec, PA=88\degr\ and 0.07\arcsec $\times$ 0.06\arcsec, PA=90\degr, respectively. The reason for the slight increase in the size of the beam for the images at 800~pc is due to the more prominent contribution of side-lobe noise at 800~pc as the source is observed to be more extended. The synthesised beam sizes and root-mean-squared (rms) noise values for both continuum and line observations are summarised in Table~\ref{t: config_specs}.

\subsection{NOEMA}

\begin{figure}
   \centering
   \includegraphics[width=0.95\hsize]{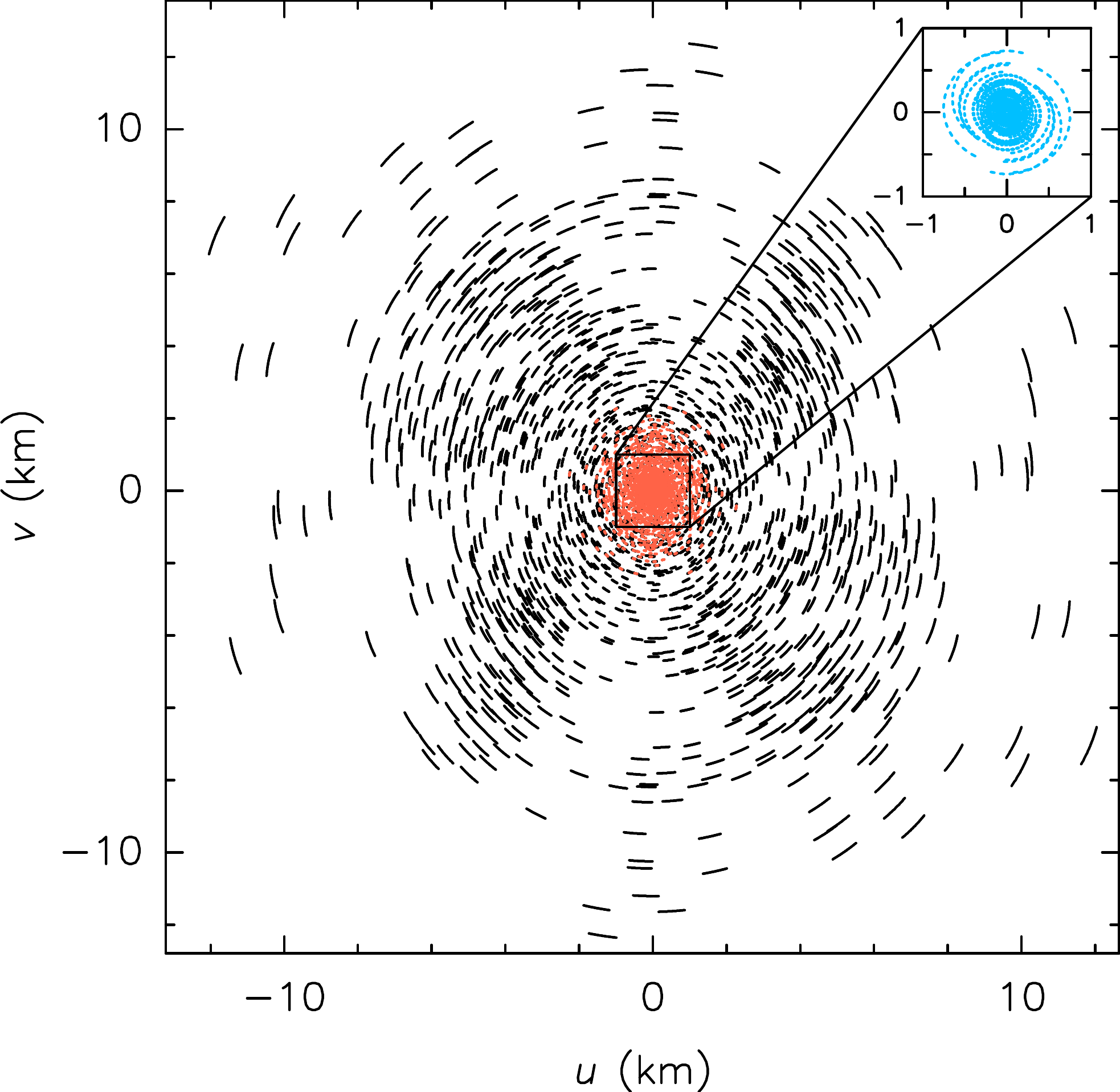}
      \caption{The \uv-coverage of simulated ALMA observations in configuration 5-9 in black and configuration 5-6 in red. The \uv-coverage of simulated NOEMA observations is shown in blue in the zoom box to the top right.}
         \label{f: uv_coverage}
   \end{figure}

For the synthetic NOEMA observations, we use the \textit{uv\_fmodel} task in the \emph{GILDAS}\footnote{\url{http://www.iram.fr/IRAMFR/GILDAS}} software package developed by IRAM and Observatoire de Grenoble. To do so, we create a \uv\ table from the input sky model by providing a \uv\ table that specifies the desired sampling of the antennas in the \uv-plane. To be able to make direct comparisons to our NOEMA large program, CORE, we use the \uv-coverage of our NOEMA observations in the A, B, and D arrays for source \W\ \citepads{2018A&A...618A..46A}. Six 15-m antennas were used at the time of the observations of \W, with baselines extending the range of 19 to 760 m. The upper right panel of Fig.~\ref{f: uv_coverage} shows the \uv-coverage of the synthetic NOEMA observations. The integration time of the synthetic observations are made in 20-minute intervals for a total time of $\sim$11 hours spread among the three different arrays. The details of the observing sequence is described in detail in \citetads{2018A&A...618A..46A}. To achieve the highest possible angular resolution, we \textit{clean}ed the synthetic visibilities using the \emph{MAPPING} program of the \emph{GILDAS} package with the \textit{clark} algorithm \citepads{1980A&A....89..377C} and a uniform weighting (robust parameter of 0.1). The images have synthesised beam sizes of 0.44\arcsec $\times$ 0.34\arcsec, PA=44\degr. The synthesised beam sizes and rms noise values for both continuum and line observations are summarised in Table~\ref{t: config_specs}.

   \begin{figure*}
   \centering
   \includegraphics[width=0.95\hsize]{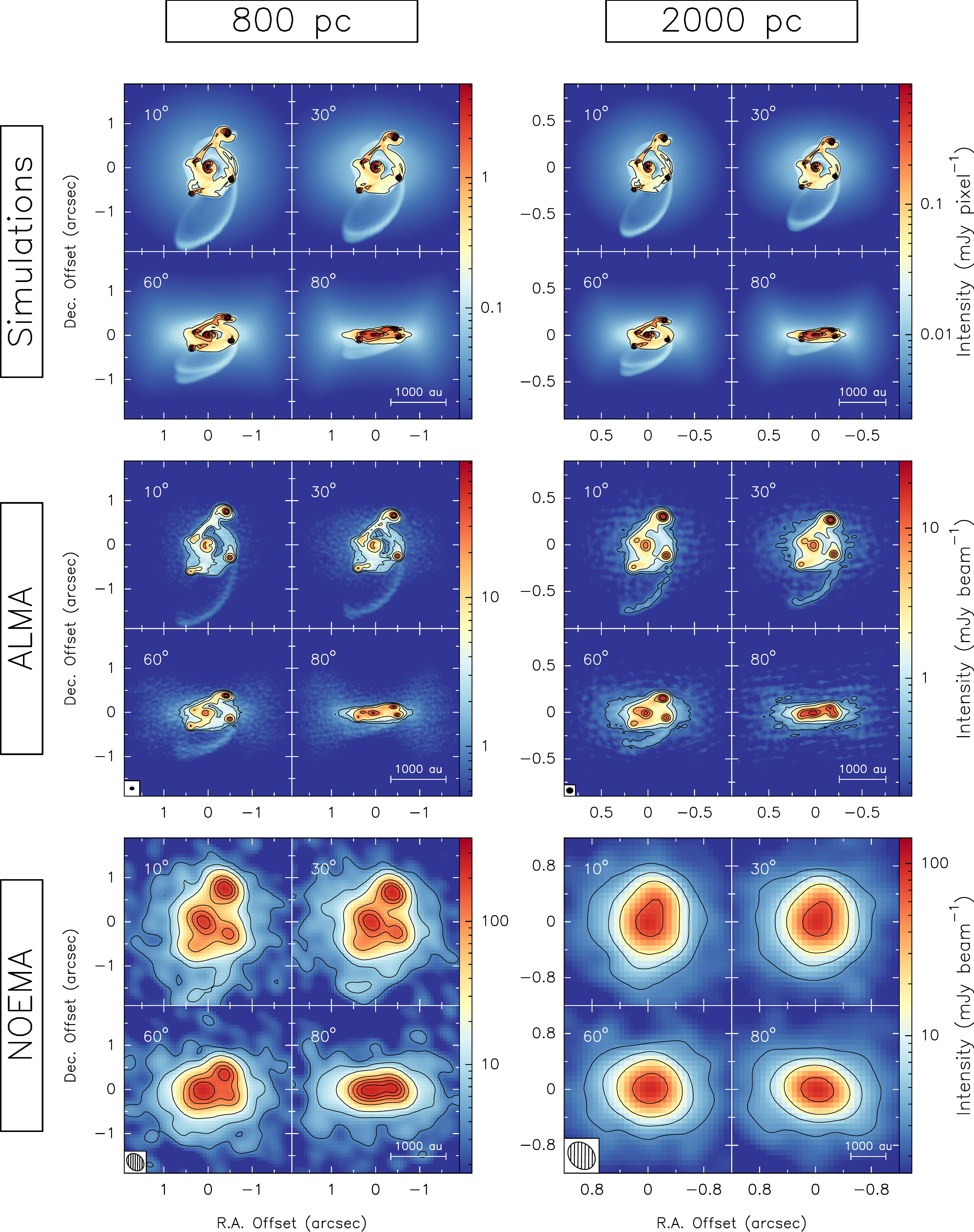}
      \caption{1.37~mm continuum images of the model simulations \emph{(top row)}, synthetic ALMA observations \emph{(middle row)}, and synthetic NOEMA observations \emph{(bottom row)} shifted to a distance of 800~pc (\emph{left column}), and 2000~pc (\emph{right column}). Each panel contains four sub-panels corresponding to the image inclined by 10$\degr$ \emph{(top left)}, 30$\degr$ \emph{(top right)}, 60$\degr$ \emph{(bottom left)}, and 80$\degr$ \emph{(bottom right)}. The contours for the model simulations are drawn at 1, 5, 9, 20, 40, and 60\% of the peak of the emission. The two outer-most contours for the synthetic observations correspond to the 6 and 12$\sigma$ levels, the rest start at 30$\sigma$ and increase in steps of 80$\sigma$ (see Table~\ref{t: config_specs} for the noise values). Synthesised beams are shown in the bottom left corners and scale bars are shown in the bottom right corners of each set of synthetic observations. Note that the intensity scale is different in each panel.}
         \label{f: continuum}
   \end{figure*}

\section{Analysis \& Results} \label{s: analysis_results}
In the following, we will present the continuum maps, and study the kinematics, physical conditions, and stability of the disks. We refer to the numerical simulations as `model simulations' and the synthetic ALMA and NOEMA observations accordingly.

\subsection{Continuum maps} \label{ss: continuum}
Figure~\ref{f: continuum} shows the 1.3~mm dust continuum models of the simulations at 800 and 2000~pc distance at four different inclinations (10\degr, 30\degr, 60\degr, and 80\degr) and the corresponding synthetic ALMA and NOEMA observations. The synthesised beam sizes, linear resolutions, and rms noise values are summarised in Table~\ref{t: config_specs}. At this snapshot, after about 12 kyr of evolution, the fragmented accretion disk spans an extent of $\sim$1000~au in dust continuum with four fragments seen on scales of $\leq$500~au, which are actively accreting material from their own smaller accretion disks (described in detail in the following section). Over-dense regions connecting the fragments are a result of gravitational instabilities in the disk which had initially formed spiral arms and resulted in the creation of the fragments. The disk at this stage is still highly dynamic, with the fragments rotating about the central protostar, migrating inwards, and encountering orbital interactions and mass transfer with each other. The center, where the protostar is actively accreting material from the disk and envelope, has a half doughnut shape as a result of the boundary conditions of the numerical simulations. The central sink has a radius of 30~au, the region within which is not included in the computation domain. As a result, the strongest continuum peak is the fragment found to the north-west of the protostar. The arc of material seen in the lower part of the disk shows the ejection of some material from the system. The halo seen around the disk corresponds to the envelope which is much less dense than the disk. 

The synthetic ALMA observations at 800~pc, corresponding to a linear resolution of 60~au, resolve all fragments at all inclinations. The rms noise of the continuum ALMA images at 800~au is higher than the case of the ALMA images at 2000~au as the extended nature of the structure results in more prominent side-lobe noise. Therefore, although the arc of material being ejected in the bottom is visible, it is not detected with 6$\sigma$ or higher certainty. On the contrary, the synthetic ALMA observations at 2000~pc have lower side-lobe noise and therefore detect this arc with 6$\sigma$ confidence. At this distance, with a linear resolution of 130~au, ALMA is able to detect all fragments but barely the closest one to the protostar and detangling the fragments at an inclination of 80\degr\ becomes difficult. This disparity in the nature of the noise and the more prominent contribution of side-lobe noise at 800~pc is the reason why the synthesised beam at 800~pc has a slightly larger size than at 2000~pc.

The synthetic NOEMA observations at 800~pc, corresponding to a linear resolution of $\sim$300~au, are able to resolve all the fragments at 10\degr\ and 30\degr\ inclinations, with only three being resolved at 60\degr\ inclination, and two at a near edge-on view of 80\degr\ inclination. At 2000~pc, corresponding to a linear resolution of $\sim$800~au, only one structure is resolved regardless of the inclination. The structure however has an elongated shape in the direction of the brightest fragment in the north-west, hinting at the possibility of unresolved fragmentation. At both distances, the disk is seen to have a larger extent due to beam smearing. The angular resolution of NOEMA is expected to improve by roughly a factor of two with the expected baseline extension in the coming years. With the upgraded NOEMA, our synthetic NOEMA observations at 2000~pc will roughly resemble the presented observations at 800~pc.

\subsection{Kinematics} \label{ss: kinematics}
In the following, we study the kinematics of the disk through moment analysis and position-velocity diagrams. 
\subsubsection{Moment maps}
We create moment maps of \mck{4} to study the disk kinematics in more detail. For the model simulations, we add arbitrary Gaussian noise to the cubes and create the moment maps with constraints on the minimum emission level set at 6$\sigma$. While there is no need to add noise to the model simulations, it makes for cleaner presentation of the data and easier comparison with maps of the synthetic observations, as well as allowing for the automation of the procedures with the observations. For the synthetic observations, due to the non-Gaussian nature of noise and the large contribution of side-lobe noise in the imaged cubes, we take a higher threshold of 18$\sigma$ to constrain our study of the kinematics to the disk region. The range of velocities over which the moment maps have been integrated is the same for the model simulations and the synthetic observations. 

\begin{figure*}
   \centering
   \includegraphics[width=0.97\hsize]{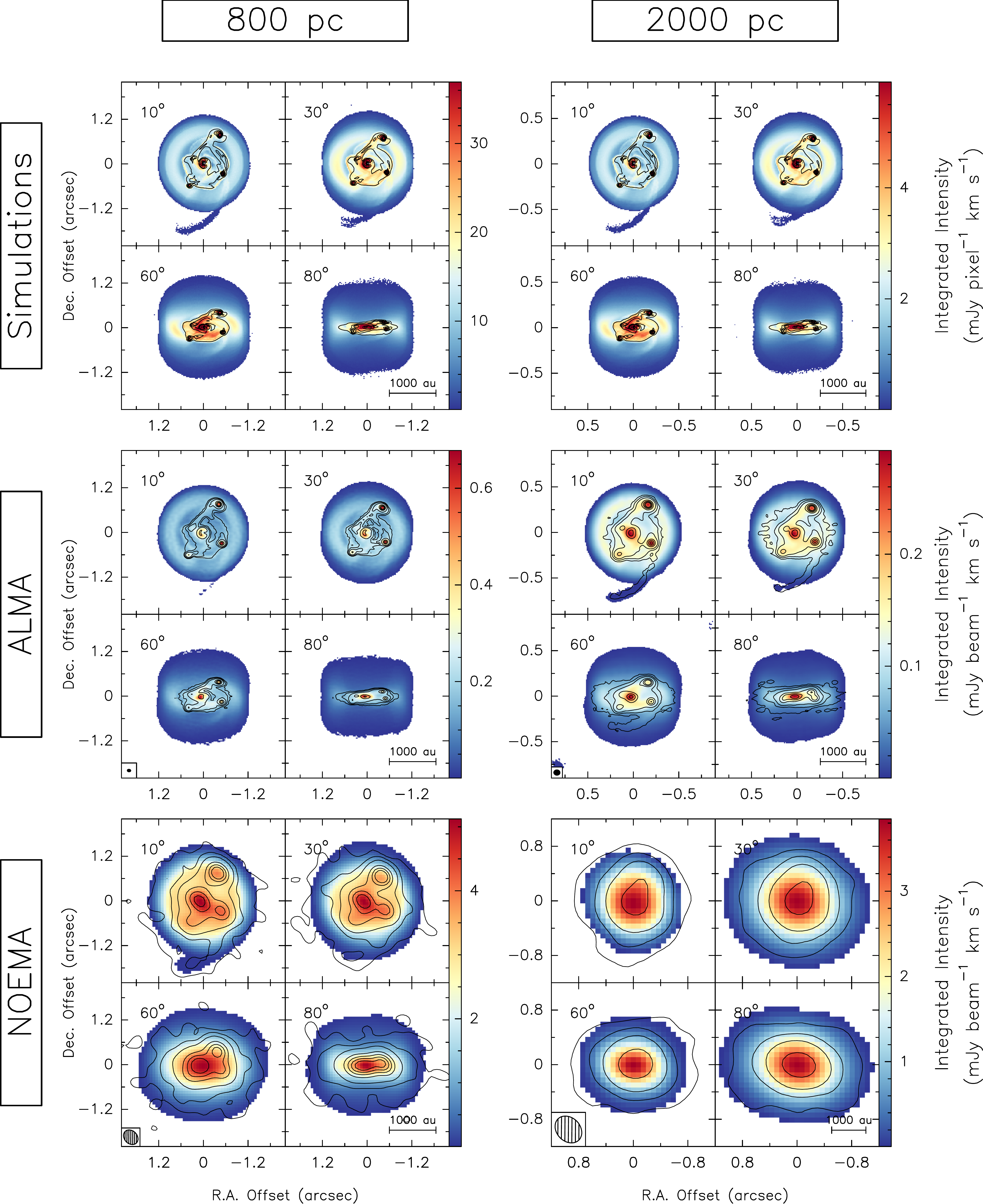}
      \caption{Integrated intensity (zeroth moment) maps of \mck{4} for the model simulations \emph{(top row)}, synthetic ALMA observations \emph{(middle row)}, and synthetic NOEMA observations \emph{(bottom row)} shifted to a distance of 800~pc (\emph{left column}), and 2000~pc (\emph{right column}). Each panel contains four sub-panels corresponding to the image inclined by 10$\degr$ \emph{(top left)}, 30$\degr$ \emph{(top right)}, 60$\degr$ \emph{(bottom left)}, and 80$\degr$ \emph{(bottom right)}. The contours correspond to 1.37~mm continuum as shown in Fig.~\ref{f: continuum}. Synthesised beams are shown in the bottom left corners and scale bars are shown in the bottom right corners of each set of synthetic observations. Note that the scale is different in each panel.}
         \label{f: ch3cn_k4_mom0}
   \end{figure*}

Figure~\ref{f: ch3cn_k4_mom0} shows the integrated intensity (zeroth moment) map of of \mck{4} for the model and synthetic observations, with continuum contour overlays. The distribution of \mc\ gas is found to be extended by a few hundred astronomical units beyond the compact dust emission, with the peaks of gas emission coinciding with dust emission peaks as expected. This extension of gas is due to the contribution of the envelope. The envelope itself is rotating and infalling onto a ring-like structure at the location of the centrifugal barrier which can best be seen in the zeroth moment maps of the 10\degr\ and 30\degr\ inclined cubes (further discussed in Sect.~\ref{ss: PV}). The arc of ejected material in the bottom is also seen in the integrated intensity maps and is well recovered in some of the synthetic observations. The distribution of gas in the synthetic NOEMA observations at 2000~pc are more symmetrically distributed than the dust emission which has an elongated shape. This is because the fraction of gas that is being accreted onto each of the fragments is completely washed out by the large beam and therefore the main contribution left is the larger-scale disk component that is feeding the protostar at the center. Similarly, in the synthetic NOEMA observations at 800~pc the peak of the integrated intensity map is at the center whereas the dust continuum peaks at the location of the fragment to the north-west, since the inner-most 30~au around the accreting protostar is not included in the computation domain, due to the existence of the sink particle, and is therefore not bright in the continuum observations. 

   \begin{figure*}
   \centering
   \includegraphics[width=0.97\hsize]{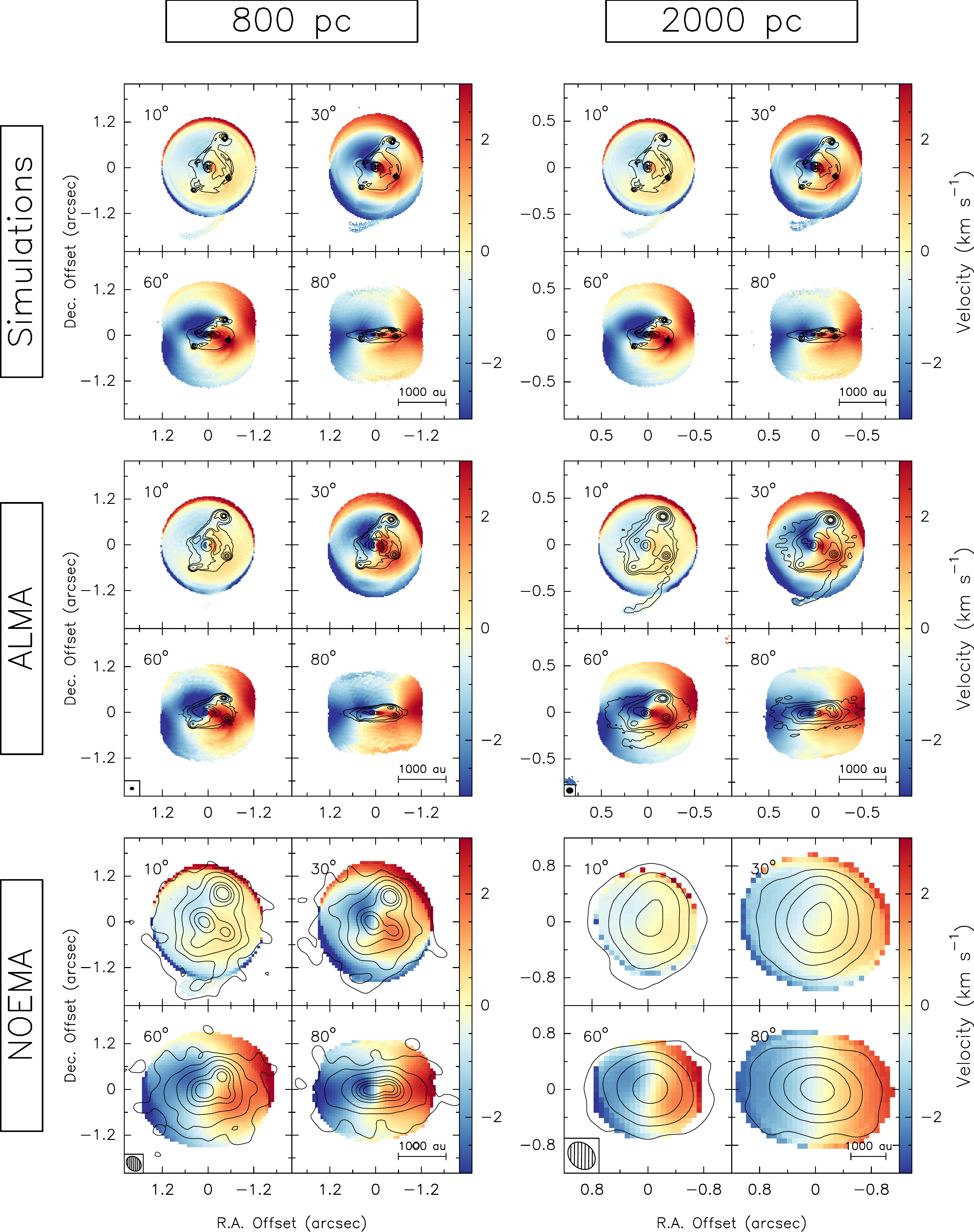}
      \caption{Intensity-weighted peak velocity (first moment) maps of \mck{4} for the model simulations \emph{(top row)}, synthetic ALMA observations \emph{(middle row)}, and synthetic NOEMA observations \emph{(bottom row)} shifted to a distance of 800~pc (\emph{left column}), and 2000~pc (\emph{right column}). Each panel contains four sub-panels corresponding to the image inclined by 10$\degr$ \emph{(top left)}, 30$\degr$ \emph{(top right)}, 60$\degr$ \emph{(bottom left)}, and 80$\degr$ \emph{(bottom right)}. The contours correspond to 1.37~mm continuum as shown in Fig.~\ref{f: continuum}. Synthesised beams are shown in the bottom left corners and scale bars are shown in the bottom right corners of each set of synthetic observations.}
         \label{f: ch3cn_k4_mom1}
   \end{figure*}

   \begin{figure*}
   \centering
   \includegraphics[width=0.97\hsize]{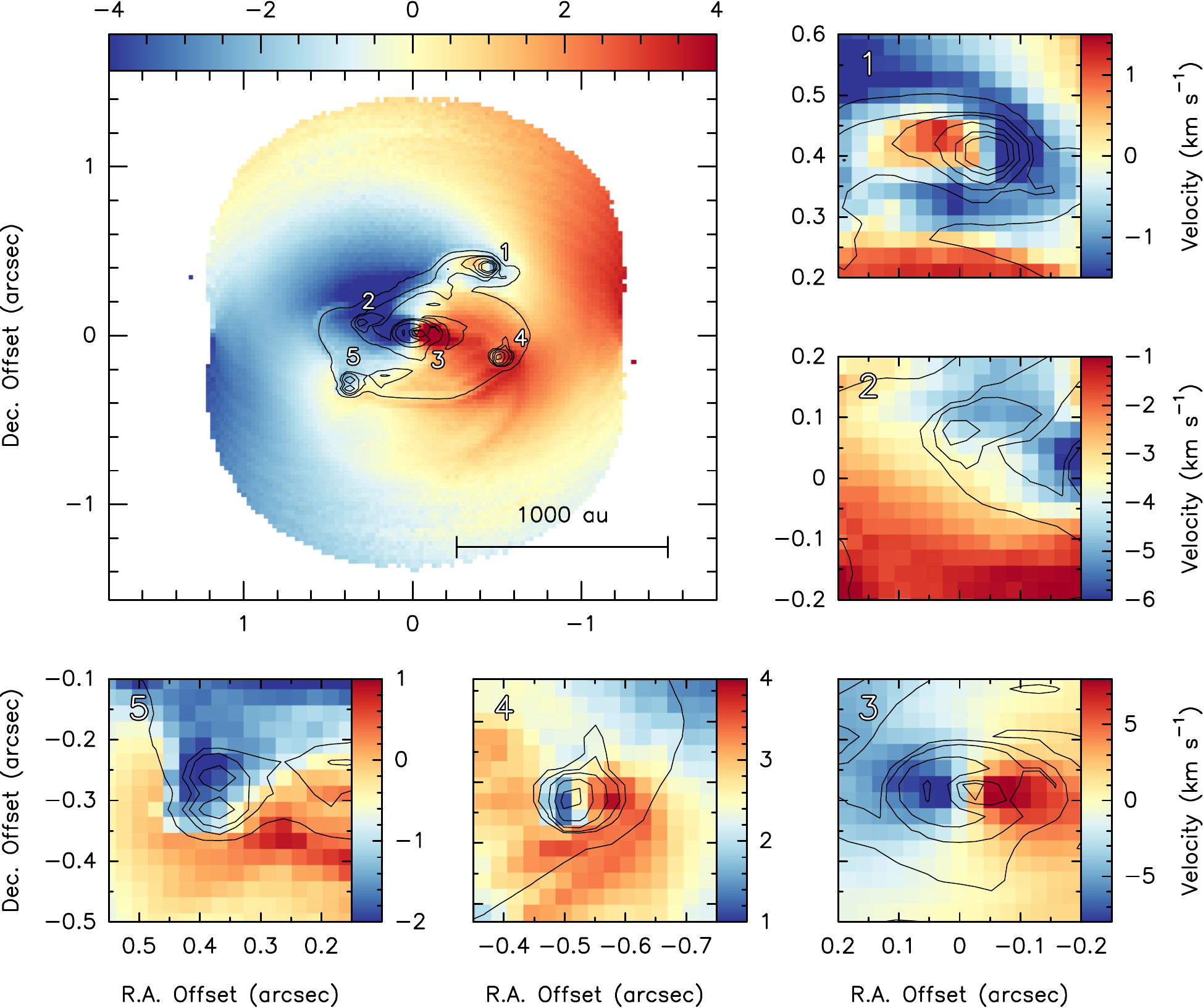}
      \caption{Intensity-weighted peak velocity (first moment) map of \mck{4} for the model simulations at an inclination of 60\degr\ and shifted to a distance of 800~pc \emph{(top left)}. The surrounding panels show zoomed views of the kinematics of each fragment as marked by the numbers in the main panel. The contours correspond to 1.37~mm continuum as shown in Fig.~\ref{f: continuum}.}
         \label{f: mom1_fragments}
   \end{figure*}

Figure~\ref{f: ch3cn_k4_mom1} shows the intensity-weighted peak velocity (first moment) map of \mck{4} for the numerical simulations and synthetic observations, with continuum contour overlays. The velocity gradient seen at 10\degr\ inclinations mainly show the contribution from the infalling and rotating envelope with the edges of the envelope shell clearly visible as it has the highest velocity offset from the local standard of rest (LSR) velocity. As the structure gets more inclined, the disk contribution becomes more clearly visible with larger velocity variation seen between the redshifted and blueshifted gas. The synthetic observations are able to recover the kinematics of the gas well, with the envelope and disk contributions becoming heavily blended in the poorly-resolved NOEMA observations making the amplitude of the velocity gradient smaller. The Yin-Yang appearance of the first moment map is due to the fragments accreting some of the infalling gas which would have otherwise accreted onto the central protostar had the fragments not existed. This can be better seen in Fig.~\ref{f: mom1_fragments} which shows zoom panels of the first moment map of \mck{4} on the positions of the fragments for the model simulations at 800~pc and inclined 60\degr. The line-of-sight velocity gradient across the central protostar is in the east-west direction and has the largest amplitude as the accretion rate is highest onto the central protostar. The rotation axis of small accretion disks surrounding each of the fragments is not necessarily inherited from the east-west rotational motion of the large-scale disk, but is set according to local dynamics. In the more inclined views towards edge-on, the Yin-Yang effect is less prominent than in the more face-on views as the fragments and their small disks are seen in one plane and start to obscure each other.

   \begin{figure*}
   \centering
   \includegraphics[width=0.92\hsize]{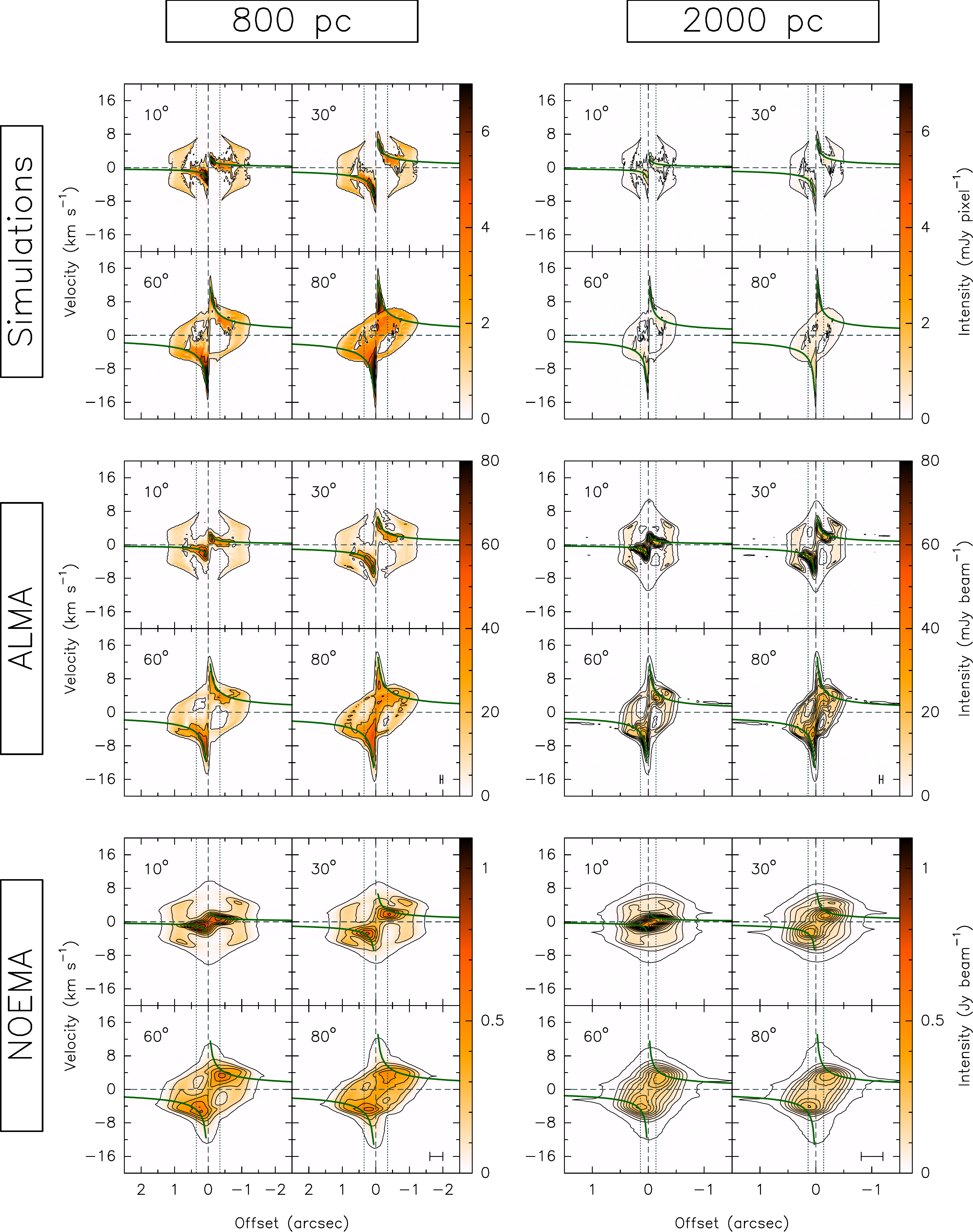}
      \caption{Position-velocity plots of \mck{4} for a cut in the east-west direction across the model simulations \emph{(top row)}, synthetic ALMA observations \emph{(middle row)}, and synthetic NOEMA observations \emph{(bottom row)} shifted to a distance of 800~pc (\emph{left column}), and 2000~pc (\emph{right column}). Each panel contains four sub-panels corresponding to the image inclined by 10$\degr$ \emph{(top left)}, 30$\degr$ \emph{(top right)}, 60$\degr$ \emph{(bottom left)}, and 80$\degr$ \emph{(bottom right)}. The three outer-most black contours correspond to the 6, 12, and 18$\sigma$ levels. The remaining black contours increase in steps of 150$\sigma$. The green curves correspond to the region within which emission is expected if the gas is in a disk in Keplerian rotation about a 10~\mo\ star. The colour scale is the same for each row to show the effect of loss of emission with increasing distance. The vertical and horizontal dashed lines correspond to the position of the protostar and zero LSR velocity. The vertical dotted lines correspond to the radii at which the centrifugal barrier is seen ($\pm250$~au). The position resolution is shown as a scale bar in the bottom right corner of each set of synthetic observations. The velocity resolution is 0.35~\kms.}
         \label{f: ch3cn_k4_pv}
   \end{figure*}

   \begin{figure*}
   \centering
   \includegraphics[width=\hsize]{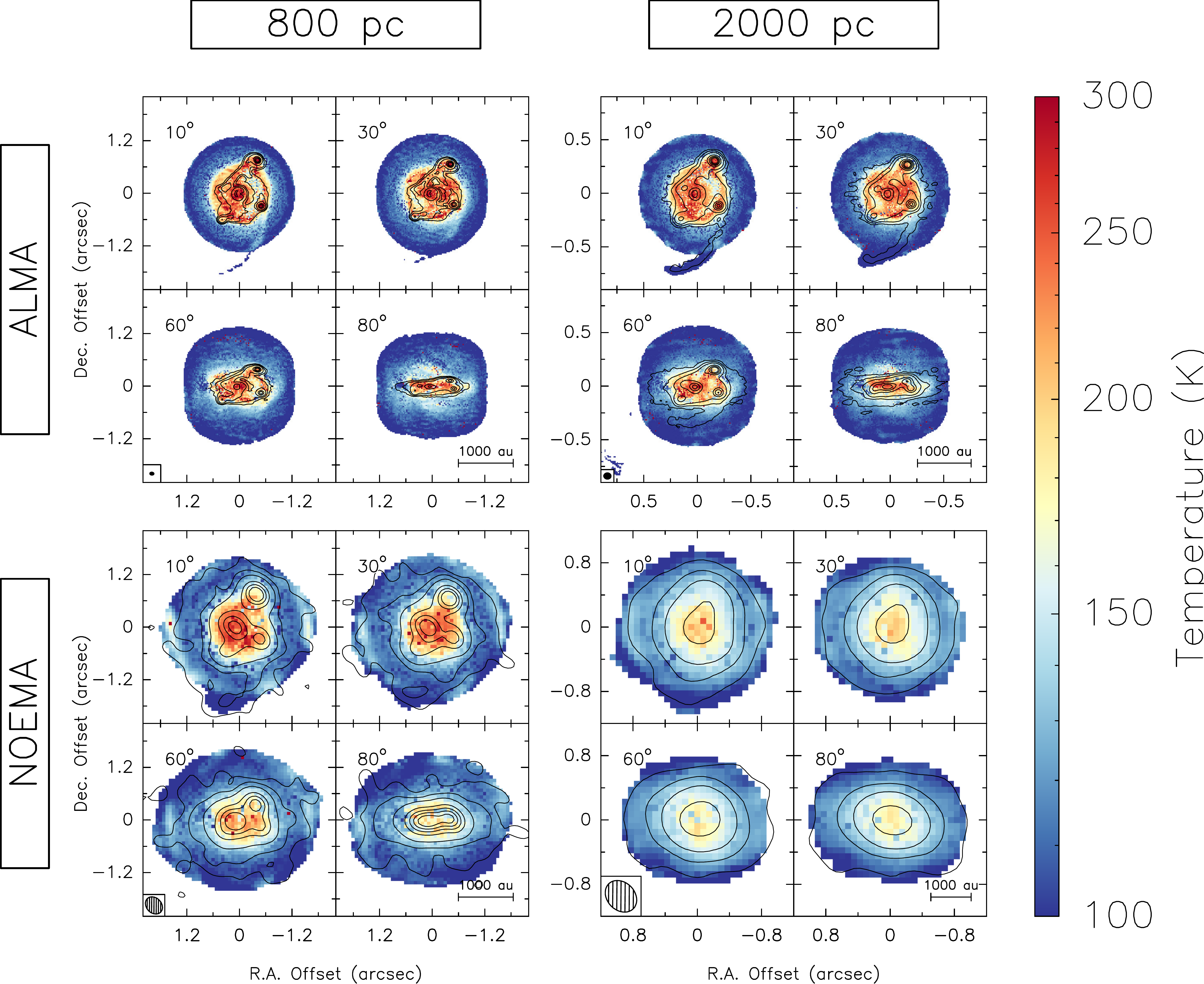}
      \caption{Rotational temperature maps of synthetic ALMA observations \emph{(top row)}, and synthetic NOEMA observations \emph{(bottom row)} shifted to a distance of 800~pc (\emph{left column}), and 2000~pc (\emph{right column}) obtained by fitting \mckr{4}{6}\ lines simultaneously with \emph{XCLASS}. Each panel contains four sub-panels corresponding to the image inclined by 10$\degr$ \emph{(top left)}, 30$\degr$ \emph{(top right)}, 60$\degr$ \emph{(bottom left)}, and 80$\degr$ \emph{(bottom right)}. Synthesised beams are shown in the bottom left corners of each set of synthetic observations. The contours correspond to 1.37~mm continuum as shown in Fig.~\ref{f: continuum}. The scale of the plots has been chosen for ease of comparison to the `true' disk mid-plane temperature as shown in Fig.~\ref{f: sim_temp}.}
         \label{f: obs_temp}
   \end{figure*}

\subsubsection{Position-velocity diagrams} \label{ss: PV}
To study the kinematics in more detail, we create position-velocity (PV) plots for a cut in the east-west direction, perpendicular to the direction of the rotation axis. Figure~\ref{f: ch3cn_k4_pv} shows the PV plots of \mck{4} for the model simulations and synthetic observation. The PV plots of the model simulations show the effect of inclination clearly in that the inner disk contribution ($r < 250$~au) is not strong in the more face-on views. As the source is more inclined towards edge-on, the disk component becomes prominent, with high-velocity contributions seen very close to the central protostar. The inner disk is in clear differential rotation with a Keplerian profile $v(r)=\sqrt{GM/r}$ about a central object with a mass of $\sim$10~\mo\ as depicted by the green curve. The contribution seen at $r > 250$~au corresponds to the region where the rotating envelope is falling onto the inner disk which has a lower radial velocity than the infall velocity of the envelope. In recent years, this transition has been observed around low-mass protostars (\eg\ \citeads{2014Natur.507...78S}; \citeads{2016ApJ...824...88O}; \citeads{2017NatAs...1E.152L}; \citeads{2017A&A...603L...3A}). These studies have found that at this position, accretion shocks lead to local heating, causing a change in the chemistry as material gets liberated off the ice mantles to the gas phase. Furthermore, evidence for the centrifugal barrier around a high-mass protostellar envelope has recently been reported by \citetads{2018A&A...617A..89C}. This transition region is estimated to be an order of magnitude closer to the central protostar in the low-mass regime (\eg\ 30--50~au for IRAS 16293--2422 Source B: \citeads{2018ApJ...854...96O}) than in the reported high-mass case (300--800~au: \citeads{2018A&A...617A..89C}). Our estimate of the centrifugal barrier located at $r\sim$250~au is in agreement with the lower end of the estimates from the observations of \citetads{2018A&A...617A..89C}. 

The sharpness of this discontinuity feature seen in the PV plots is a function of resolution and inclination angle. It can be seen in the NOEMA observations at 2000~au that the `kink' seen in the other PV plots as a result of this transition is completely smoothed out. Furthermore, the velocity profile of the infalling rotating envelope would look different depending on the offset between the line-of-sight and the source position as depicted in Figure~3b of \citetads{2014Natur.507...78S}, which can have an effect on the sharpness of this discontinuity. The PV plots presented in this work have been made for a cut that goes through the protostar as we have knowledge of the precise location of the protostar.  

The PV plots of the synthetic ALMA observations resemble the PV plots of the model simulations but smoothed in the position direction as expected. The high-velocity components are still clearly detected in the more inclined cases, but the spreading of emission would mean that the Keplerian curve corresponding to rotation of matter about a $\sim$10~\mo\ star does not perfectly fit anymore, especially in the 2000~pc case. As we move to the synthetic NOEMA observations and the emission gets even more smeared in the position direction, the high-velocity components seen in the more inclined views become hard to detect and the Keplerian curve does not fit well to the observations. This is best seen in the synthetic NOEMA observations at 2000~pc where the inner disk and envelope contributions are completely blended and the resulting PV plots better resemble rigid-body rotation than differential rotation, similar to many observational data in the last decade at comparable resolution.

As temperature is expected to be higher in regions of the disk closer to the protostar, one would expect line transitions which require higher excitation energies to be present in regions closer to the star. One signpost of differential rotation in observations is often seen in the steepening of the slope of the emission in the PV diagrams for the higher excited lines, such that a transition that is tracing the innermost regions of the disk would have higher velocities at smaller radii than the transitions that trace the entire disk region. This has been observed in the case of higher-$K$ transitions of \mc\ \citepads{2015ApJ...813L..19J}, or vibrationally excited \mc\ transitions \citepads{2017A&A...602A..59C}. Comparing the PV plots of \mck{2}\ to \mck{4}\ and \mck{6}, we do not see this steepening of the slope with higher-$K$ levels because all lines trace regions with similar optical depths and therefore probe similar regions. 

\subsection{Temperature distribution} \label{ss: temperature}

To gain a better understanding of the disk structure, we calculate the rotational temperature and column density of the dense-gas tracer \mc\ under the assumption of Local Thermodynamical Equilibrium (LTE). To do so, we use the eXtended \emph{CASA} Line Analysis Software Suite \citepads[\emph{XCLASS\footnote{\url{https://xclass.astro.uni-koeln.de}}},][]{2017A&A...598A...7M} software in \emph{CASA}. In Fig.~\ref{f: obs_temp} we show the resulting rotational temperature maps obtained from pixel-by-pixel \emph{XCLASS} modelling of \mckr{4}{6} for the synthetic observations. The recipe is as outlined in Appendix~B of \citetads{2018A&A...618A..46A}, where we discuss the reasoning behind not including the lower $K$-ladder transitions in the fitting routine. In these runs, we have also allowed for the size of the source to be a fitting parameter rather than fixing it to a value larger than the beam. This strategy has been tested and yields smoother temperature maps. For each synthetic observation, we calculate the rms noise in the emission-free region in the channel that has the peak of emission for \mck{4}, and only fit pixels with signal higher than 6 times this noise. We note that the spectra at the positions where the envelope is directly falling onto the disk are triply peaked corresponding to red- and blue-shifted infall from the envelope onto the disk with a third component corresponding to the rotational motion of the disk itself. This is particularly prominent in the views closer to face-on. Since performing a three-component fit to the entire map was not feasible, we smoothed the velocity resolution by a factor of 8, reaching a spectral resolution of 2.8~\kms, to blend all three components of the lines enough to perform a one-component fit to the spectrally smoothed data cubes. 

\begin{figure}
   \centering
   \includegraphics[width=\hsize]{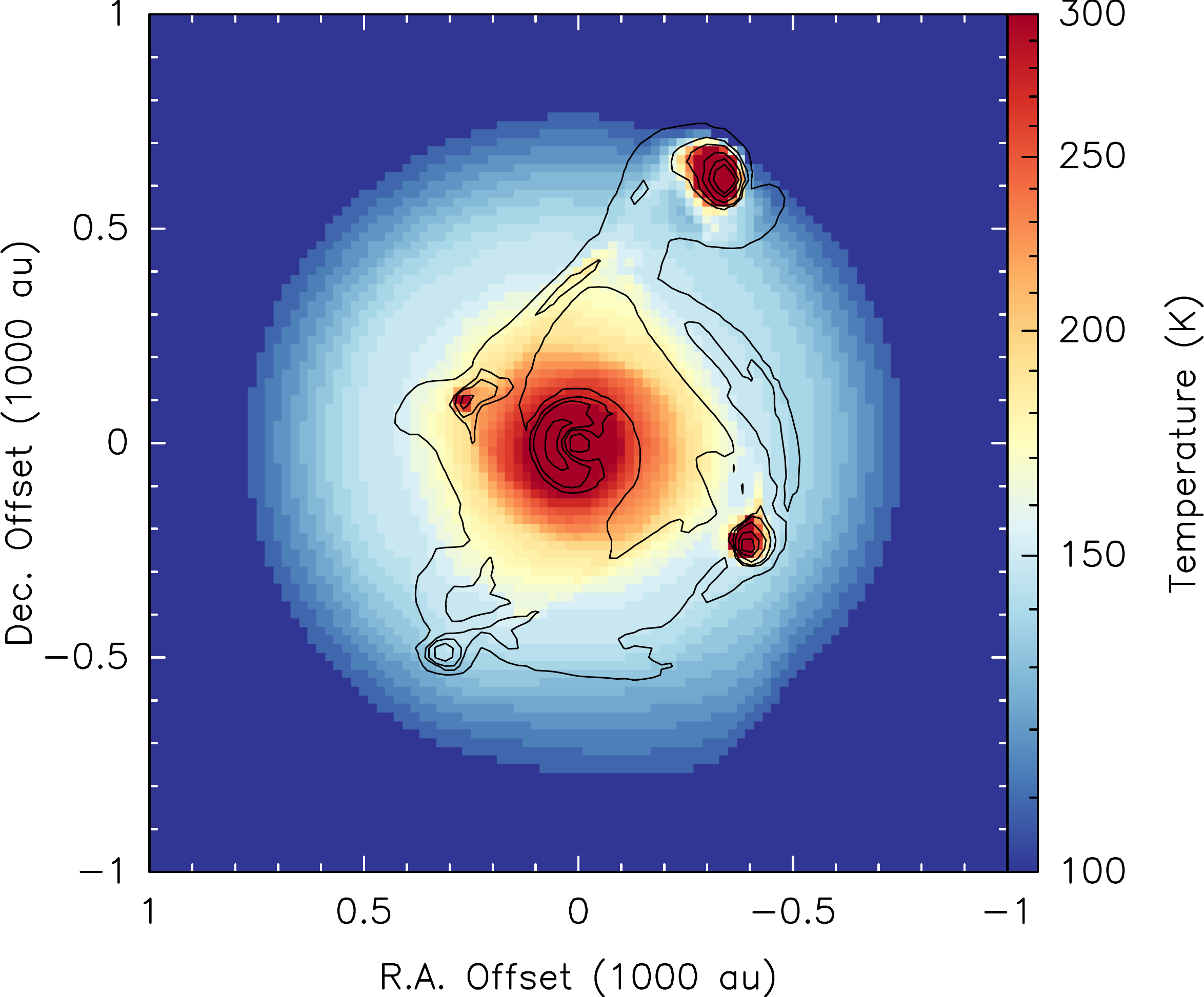}
      \caption{Face-on view of the disk mid-plane temperature from the model simulations. The contours correspond to the continuum image and are drawn at 1, 5, 9, 20, 40, and 60\% of the peak of the emission.The scale of the figure has been chosen for ease of comparison to the synthetic observations as shown in Fig.~\ref{f: obs_temp}. Temperatures at the locations of the fragments get as high as 1000~K.}
         \label{f: sim_temp}
   \end{figure}

Figure~\ref{f: obs_temp} shows the resulting rotational temperature maps for the synthetic observations. Under the LTE assumption, the rotational temperature of \mc\ can be accepted as the kinetic temperature of the gas. The median gas temperatures obtained from the synthetic observations at each inclination are listed in Table~\ref{t: temp_mass_Toomre}. On average, temperature decreases as the system becomes more inclined, which is expected as the observed column densities are higher in the more inclined views. Also, the arc of material being ejected from the system seen to the bottom of the disk has a higher temperature than its surroundings. For comparison, the face-on view of the disk mid-plane temperature from the model simulations is presented in Fig.~\ref{f: sim_temp} on the same logarithmic scale. Mid-plane temperatures at the location of fragments can reach values higher than 1000~K, but on average the median temperature for the inner disk ($r< 400$~au) is 177~K, about 50~K warmer than the outer envelope ($400~\mathrm{au}<r<800~\mathrm{au}$), which has a median temperature of 128~K. In Table~\ref{t: temp_mass_Toomre}, we also list median temperature values calculated for the regions within 12$\sigma$ dust continuum contours (the second outermost contours in Fig.~\ref{f: obs_temp}) which better correspond to the disk. On average, the temperatures are as expected higher in the disk than in the outer envelope. Moreover, the temperature decreases as angular resolution worsens when the disk structure is no longer recovered. Interestingly, the temperature distributions for the synthetic NOEMA observations at 2000~pc at 10\degr\ and 30\degr\ inclinations are elongated in the same direction as the continuum, while this elongation feature is not seen in the respective zeroth moment maps of \mck{4} (bottom right panel of Fig.~\ref{f: ch3cn_k4_mom0}).

\subsection{Mass estimates} \label{ss: masses}
At the snapshot of the numerical simulations under investigation, the protostar has gained a mass of $\sim$10~\mo. In the following, we use different techniques often used in observations for determining core and protostellar masses. 

\begin{table*}[!ht]
\renewcommand{\arraystretch}{1.3}
\caption{Summary of temperatures, masses, and $Q$ values for the synthetic observations.}
\centering
\label{t: temp_mass_Toomre}
\begin{tabular}{lccccccc}
\hline\hline
Interferometer & Distance  & Inclination &  Median $T$\tablefootmark{a} & Median $T_{12\sigma}$\tablefootmark{b} & Mass\tablefootmark{c} & Median $Q$ \\
 & (pc) & (degrees) &  (K) & (K) & (\mo)&    \\
\hline
\multirow{4}{*}{ALMA} & 800 & 10 & 115 & 228 & 2.9 & 2.6 \\
 &  800 & 30 & 112 & 213 & 3.2 & 2.3 \\
 &  800 & 60 & 110 & 204 & 3.3 & 1.5 \\
 &  800 & 80 & 110 & 202 & 2.8 & 0.5 \\
 \hline
 \multirow{4}{*}{ALMA} & 2000 & 10 & 112 & 214 & 2.9 & 3.5 \\
 &  2000 & 30 & 111 & 211 & 3.0 & 3.1 \\
 &  2000 & 60 & 108 & 187 & 3.0 & 2.0 \\
 &  2000 & 80 & 108 & 173 & 2.5 & 0.7 \\

\hline
\multirow{4}{*}{NOEMA} & 800 & 10 & 121 & 153 & 2.6 & 5.2 \\
 & 800 & 30 & 118 & 145 & 2.7 & 4.7 \\
 & 800 & 60 & 115 & 142 & 2.6 & 3.1 \\
 & 800 & 80 &114 & 139 & 2.4 & 1.0 \\
\hline

\multirow{4}{*}{NOEMA} & 2000 & 10 & 125 & 148 & 3.5 & 3.6 \\
 & 2000 & 30 & 125 & 146 & 3.4 & 3.4 \\ 
 & 2000 & 60 & 123 & 143 & 3.3 & 2.0 \\
 & 2000 & 80 & 119 & 143 & 2.9 & 0.5 \\
 \hline
\end{tabular}
\tablefoot{
\tablefoottext{a}{Median temperature, calculated from the maps presented in Fig.~\ref{f: obs_temp}.}
\tablefoottext{b}{Median temperature within 12$\sigma$ dust continuum  (second outermost) contours as presented in Fig.~\ref{f: obs_temp}.}
\tablefoottext{c}{Mass is calculated from the dust emission using Eq.~\ref{e: mass}.}
}
\end{table*}

\subsubsection{Masses from dust emission} \label{ss: core_mass}
Because dust opacity decreases with increasing wavelength, thermal emission of dust at longer wavelengths in the Rayleigh-Jeans limit can trace large column densities and provide estimates for the dust mass if the dust emissivity is known \citepads{1983QJRAS..24..267H}. Assuming optically thin dust emission at 1.3~mm, for a gas-to-dust mass ratio $R$, and a dust absorption coefficient $\kappa_\nu$, we can convert the flux density $F_\nu$ of the continuum observations to a mass via
\begin{equation}
  \label{e: mass}
  M=\frac{d^2\,F_\nu\,R}{B_\nu (T_D)\,\kappa_{\nu}},
\end{equation}
where $d$ is the distance to the source, and $B_\nu (T_D)$ is the Planck function which at the wavelengths under investigation follows the Rayleigh-Jeans law and is linearly dependent on the dust temperature $T_D$. We adopt a value of 150 for the gas-to-dust mass ratio \citepads{2011piim.book.....D} and $\kappa_\nu=0.9$~$\mathrm{cm^2\,g^{-1}}$ corresponding to thin ice mantles after $10^5$ years of coagulation at a density of $10^6$~cm$^{-3}$  \citepads{1994A&A...291..943O}. We assume the kinetic temperature of the gas, derived from the radiative transfer modelling of \mckr{4}{6} as presented in Sect.~\ref{ss: temperature}, to be coupled to the dust temperature. Therefore, we use the temperature maps presented in Fig.~\ref{f: obs_temp} together with the continuum maps shown in Fig.~\ref{f: continuum} to create mass maps. Summing over all pixels, we then obtain a total mass for each synthetic observation, summarised in Table~\ref{t: temp_mass_Toomre}. The values are on average between 2.5--3.5~\mo. For comparison, the total mass within the inner 8400 by 8400~au box for the model simulations is $\sim$14~\mo\ calculated in the same manner by using the face-on continuum map and the mid-plane temperature shown in Fig.~\ref{f: sim_temp}, which accounts for the total gas mass in this area. The bulk of the remaining mass is on larger scales not included in our analysis. Therefore, the mass estimates from the synthetic observations recover roughly 20\% of the total mass as calculated from the model simulations and the bulk of the flux is filtered out by the interferometer. 

Considering only the synthetic ALMA observations, the mass estimates at 2000~pc are on average smaller than at 800~pc, by roughly 10\%. Since the temperature distribution on average does not vary significantly between the two cases, one would expect the mass estimate to also stay consistent. This small excess of mass corresponds to the contribution from the halo of material found outside of the 6$\sigma$ continuum contours in the middle panels of Fig.~\ref{f: continuum}. Since the continuum maps at 800~pc are noisier than at 2000~pc (see Table~\ref{t: config_specs}) the mass estimates at 800~pc are higher than at 2000~pc. Conversely, the mass estimates at 2000~pc for the synthetic NOEMA observations are roughly 20\% higher than at 800~pc. While the \textit{average} temperature is not significantly different in the two cases (see Table~\ref{t: temp_mass_Toomre}), the distribution of temperature in the inner-most regions where the bulk of continuum emission is concentrated is significantly cooler in the 2000~pc case than at 800~pc (see bottom row of Fig.~\ref{f: obs_temp}), resulting in a higher mass estimate at the farther distance. 

\subsubsection{Protostellar masses from PV plots} \label{ss: pv_mass}
Protostellar masses are often estimated from fitting Keplerian profiles with $v(r)=\sqrt{GM/r}$ to PV diagrams. In Fig.~\ref{f: ch3cn_k4_pv} we show the PV plots for the model simulations and synthetic observations for a cut along the east-west direction going through the center of the sink particle where the protostar is located. While the green curves correspond to the region within which emission is expected if the gas is in a disk in Keplerian rotation about a 10~\mo\ star, they do not represents fits to the PV plots. In order to estimate protostellar masses from these PV plots, we use the \emph{KeplerFit} package\footnote{Developed by Felix Bosco, available at \url{https://github.com/felixbosco/KeplerFit}.} (described in \citeads{2019A&A...629A..10B}) which uses the method presented in \citetads{2016MNRAS.459.1892S}. The method first identifies the two quadrants of the PV plot that contain most of the emission, and then starting from the positions closest to the protostar moving outwards it determines the highest velocity at which the signal becomes higher than a given threshold. This then determines the boundary which these authors call the `upper edge' of the PV plot which is then fit with a Keplerian profile to determine the enclosed mass. This method has proven to yield more accurate mass estimates than fitting the pixels with the maximum intensity for each radius or for each velocity channel \citepads{2016MNRAS.459.1892S}. 

\begin{figure}[t!]
   \centering
   \includegraphics[width=\hsize]{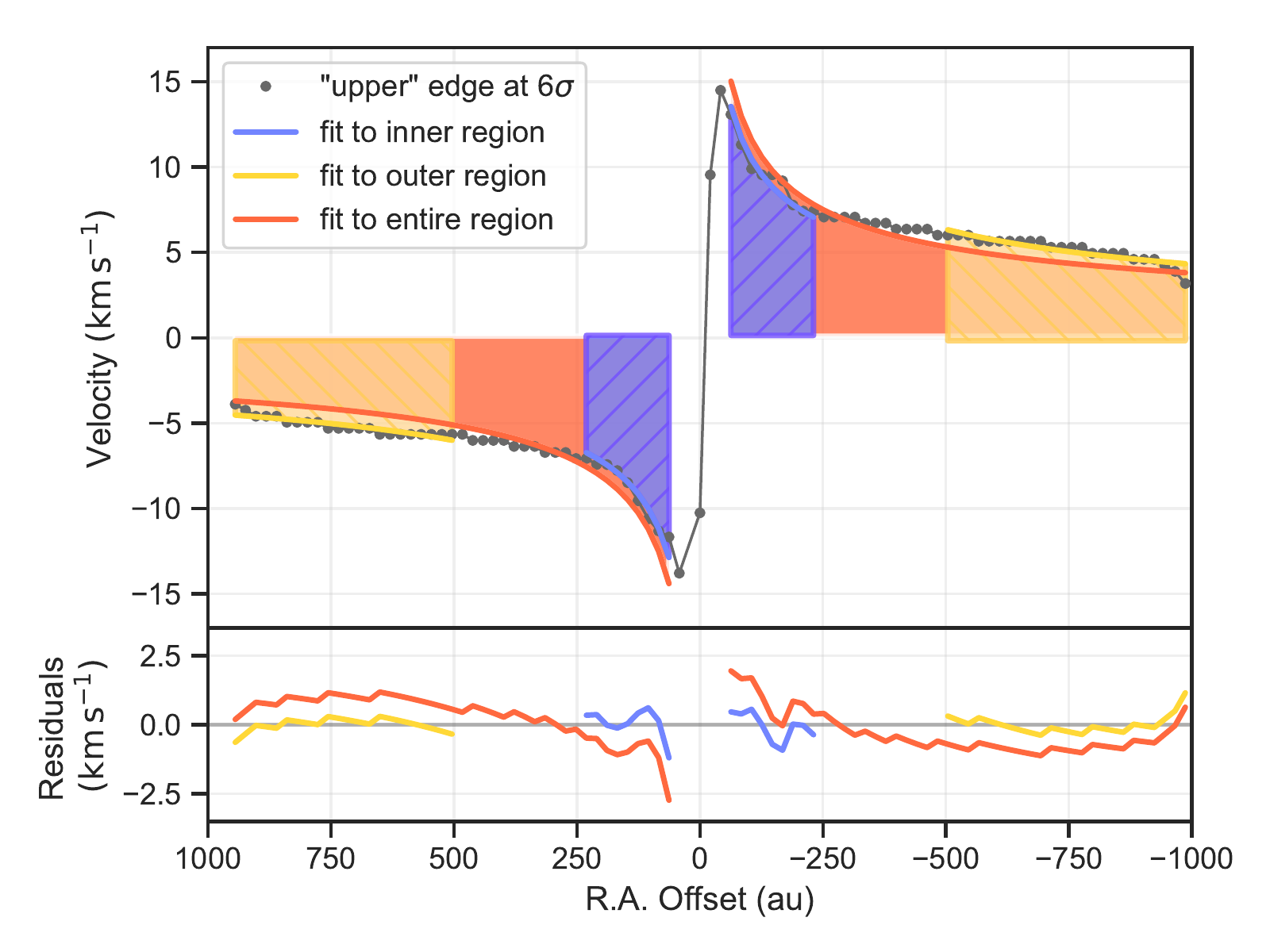}
      \caption{An example of the Keplerian fitting approach based on \citetads{2016MNRAS.459.1892S} to the 6$\sigma$ outer edge of the PV plot of \mck{4} for the model simulation inclined to 60\degr\ at a distance of 800~pc in order to determine the protostellar mass. The fits to the inner, outer, and entire regions yield protostellar mass estimates of 12.4, 21.4, and 15.4~\mo, respectively. The inner region traces the disk which follows a Keplerian rotation curve better than the outer region which has contributions from the rotating and infalling envelope. The bottom panel shows the residuals. The protostar has a mass of $\sim$10~\mo\ while the disk has a mass of $\sim$8~\mo.}
         \label{f: pv_fit}
   \end{figure}

\begin{figure*}[th!]
   \centering
   \includegraphics[width=0.5\hsize]{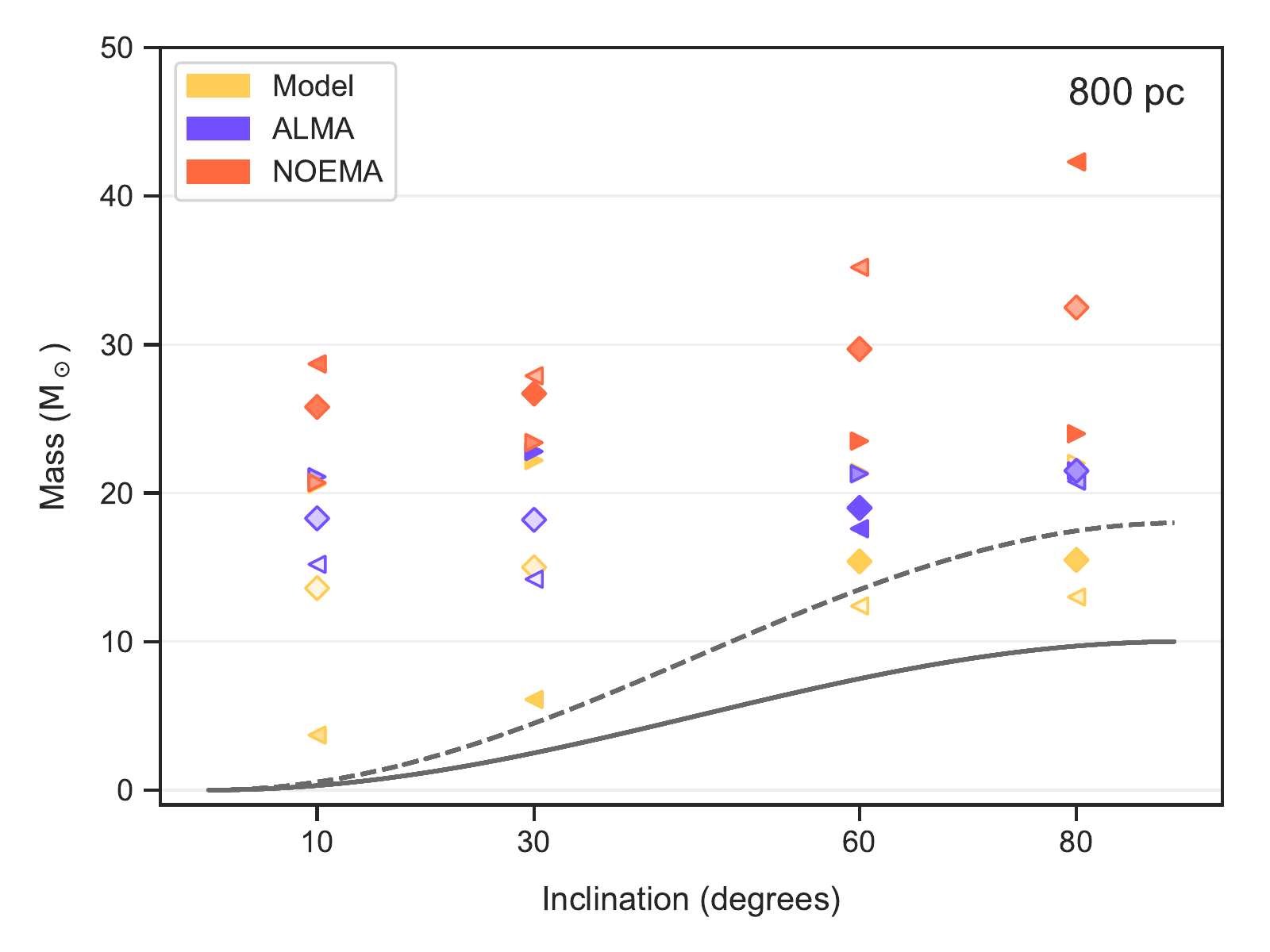}\includegraphics[width=0.5\hsize]{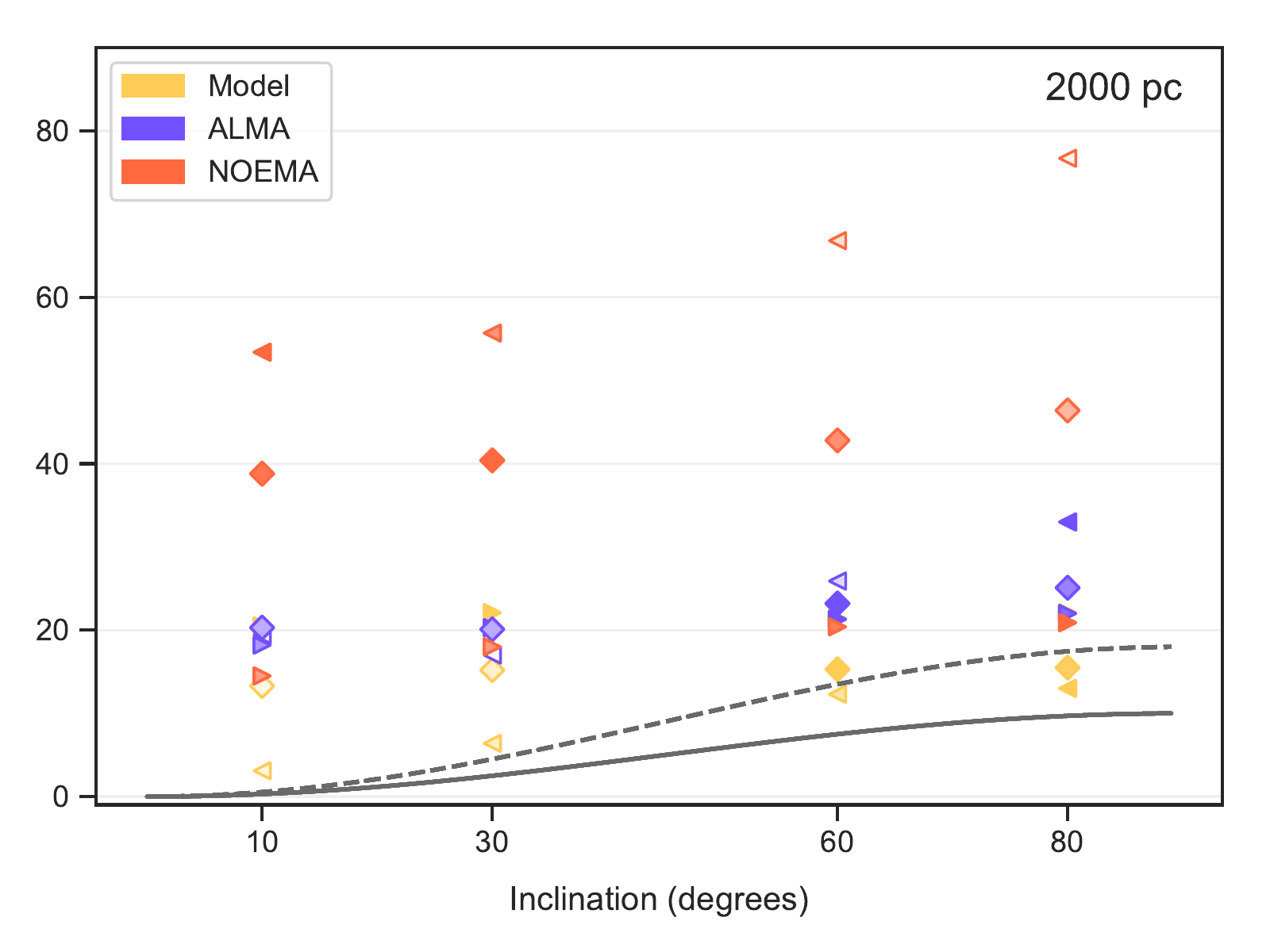}
      \caption{Mass estimates from fitting the 6$\sigma$ edges of the PV diagrams of \mck{4} shown in Fig.~\ref{f: ch3cn_k4_pv} using the method introduced by \citetads{2016MNRAS.459.1892S} for the model simulations \emph{(yellow)}, ALMA \emph{(purple)}, and NOEMA observations \emph{(orange)} at 800~pc \emph{(left)} and 2000~pc \emph{(right)}. The triangles pointing to the left and right correspond to fits to the inner and outer regions, respectively. The diamonds correspond to fits to the entire PV curves. The transparency of the markers correspond to the $\chi^2$ value of the fit, weighted by the $\chi^2$ value of the best-fit curve for a given region. The more opaque the marker, the better the fit. The solid and dashed grey curves correspond to the true estimate for the mass of the protostar (10~\mo), and the mass of the protostar + disk (18~\mo), respectively, corrected by sin$^2$($i$) for each inclination $i$.}
         \label{f: pv_fit_all}
   \end{figure*}

It can clearly be seen from the PV plots shown in Fig.~\ref{f: ch3cn_k4_pv} that the rotating and infalling envelope (regions outside of the vertical dashed lines) shows different kinematic signatures than the inner disk as discussed in Sect.~\ref{ss: PV}. More specifically, there exists a discontinuity in the resolved PV plots as a result of the infalling and rotating envelope having a higher infall velocity than the inner disk. Therefore fits to the inner regions would be expected to yield lower mass estimates than to the outer envelope which at a given position has higher velocities. For this reason, we fit the inner and outer part of the disk separately and compare the results to the case where the entire PV diagram is fitted. An example is shown in Fig.~\ref{f: pv_fit} for the model simulations at 800~pc. We always fit the 6$\sigma$ outer edge for a range of radii and always exclude in the fitting the regions closest to the protostar where the velocity decreases with decreasing distance to the protostar as a result of finite resolution (see Appendix A.3.2 of \citeads{2019A&A...629A..10B}). The resulting mass estimates from fitting the inner, outer, and entire PV curves for the model simulations and synthetic observations are presented in Fig.~\ref{f: pv_fit_all}. Knowing that the central protostar has gained a mass of $\sim$10~\mo\ and that the disk mass is $\sim$8~\mo\ at this snapshot of the simulation, we can test the accuracy of estimating the enclosed masses using this method in each scenario. We will first focus on the general findings.

As rotational velocity is expected to scale with inclination angle $i$ by a factor of $\mathrm{sin}(i)$, the mass estimates would be expected to scale by a factor of $\mathrm{sin}^2(i)$. Therefore, it is expected that fitting the kinematics of the more inclined disks would result in higher mass estimates than their less inclined counterparts (grey curves in Fig.~\ref{f: pv_fit_all}). While this general trend is seen in our fitting results, we find that the fitted masses are on average higher than the theoretical expectation, similar to the findings of \citetads{2016MNRAS.459.1892S}. For example, one would expect the enclosed mass estimate at 30\degr\ to be 0.25 times lower than the true value of 18~\mo\ at 90\degr\ inclination; however, the mass estimate from the fitting the inner region of the model simulation is larger than the expected 4.5~\mo\ by 35\%. \citetads{2016MNRAS.459.1892S} attribute this increase to the existence of considerable radial motions which in the more inclined views noticeably contribute to the line-of-sight velocities. It is worth noting that while the 10~\mo\ protostar is located at the center of the computational domain at zero offset in the PV diagrams, the disk which roughly contains 8~\mo\ spans a range of radii, such that the mass-velocity relation is $v(r)=\sqrt{G(M_\ast+M_\mathrm{disk}(r)/r}$. Therefore, the enclosed mass is not exactly 18~\mo\ at all radii and the true theoretical curve lies somewhere between the solid and dashed lines in Fig.~\ref{f: pv_fit_all}. Interestingly, the fits to the inner disk of the model simulations inclined to 60\degr\ and 80\degr\ provide mass estimates in this range. 

For the model simulations where we fit the PV curves of the model simulations and the ALMA observations at 800~pc, the protostellar mass estimates obtained from fitting the inner disk regions are as expected smaller than the estimates from fitting the outer or entire regions, and provide estimates closest to the expected values. For these cases, fits to the entire region yield masses that are less than the mass estimates obtained from fitting the outer regions. In these cases, fits to the outer regions overestimate the masses because the component that is fitted is mainly the envelope within which more mass is contained and which does not necessarily have a Keplerian-like rotation profile. 

For all observations, the method overestimates the mass, even when fitting the inner disk regions. The mass is much more overestimated for disks with lower inclinations (i.e., views closer to face-on). The fact that we generally overestimate the masses especially at lower inclinations can be seen as favourable since without the knowledge of the inclination in most observations, one typically adopts the mass estimate obtained from fitting the PV diagram as the protostellar mass without any corrections for inclination. Therefore, it is somewhat reassuring that fits to the entire region for the model simulations and all ALMA observations, which properly resolve the disk, are quite close to the expected enclosed mass of $\sim$18~\mo\ at all inclinations. 

As resolution worsens, in the case of the synthetic ALMA observations at 2000~pc and synthetic NOEMA observations at both distances, the fits to the outer regions provide the lowest mass estimates which are also closer to the expected value. For the ALMA observations, the variations in the mass estimates from fits to different regions is not as significant as for the NOEMA observations at 800 and 2000~pc where the fits to the inner regions can yield masses that are larger than the fits to the outer regions by almost a factor of 2 and 4, respectively. 

While the \citetads{2016MNRAS.459.1892S} method of fitting the outer edge of the PV diagram for determining the enclosed mass is the most accurate method currently used, we find that fitting the entire PV diagram of poorly-resolved observations yields masses that are highly overestimated. This is expected as the emission is smeared over a larger area with the envelope and disk components completely blended, resulting in the PV diagrams not having a Keplerian-like shape  (see bottom row of Fig.~\ref{f: ch3cn_k4_pv}). In such cases, we find that fitting the outer regions of the PV diagram (in our cases, regions beyond 1600~au) provides better estimates for the protostellar mass. The same analysis was done for the PV plots of \mck{K}, $K=3, 5, 6$ lines and the variations in the mass estimates were marginal as compared with those presented in Fig~\ref{f: pv_fit_all} for the $K=4$ transition. The only noticeable difference was that the mass estimates obtained from fitting the PV diagrams of the $K=3$ transition were slightly higher than those obtained from the other lines, as this transition is more easily excited than the others and the emission is hence distributed over a larger area.

\subsection{Toomre stability} \label{ss: toomre}

As mentioned in the introduction, the \tq\ parameter can be used to study the stability of a disk against gravitational collapse \citepads{1964ApJ...139.1217T}. The stabilising force of gas pressure is included in the expression of $Q$ (Eq.~\ref{e: Toomre}) via the sound speed,
\begin{equation}
  c_s=\sqrt{\frac{\gamma k_\mathrm{B} T}{\mu m_\mathrm{H}}},
\end{equation}
where $\gamma$ is the adiabatic index with a value of $7/5$ for diatomic gas,  $k_\mathrm{B}$ is the Boltzmann constant, $\mu$ is the mean molecular weight with a value of 2.8, and $m_\mathrm{H}$ is the mass of the hydrogen atom. Furthermore, we account for the stabilising force of shear by assuming the disk is in Keplerian-like rotation such that the angular velocity of the disk at a given radius $r$ would be 
\begin{equation} \label{e: omega}
\Omega(r)=\sqrt{\frac{G\,M}{r^3}}.
\end{equation}

\begin{figure}
   \centering
   \includegraphics[width=\hsize]{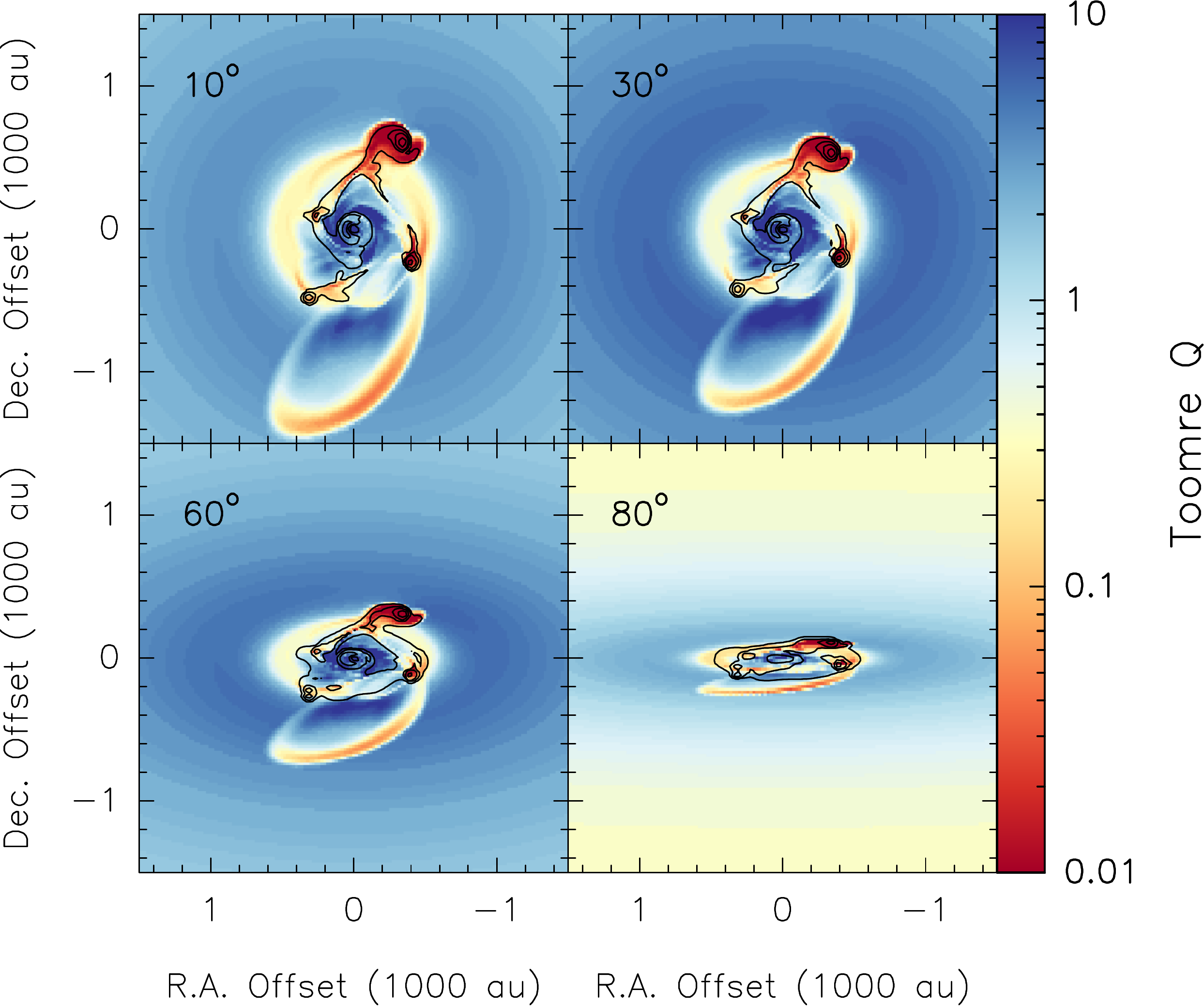}
      \caption{The `true' $Q$ map obtained from the model simulations inclined by 10$\degr$ \emph{(top left)}, 30$\degr$ \emph{(top right)}, 60$\degr$ \emph{(bottom left)}, and 80$\degr$ \emph{(bottom right)}. The contours correspond to 1.37~mm continuum as shown in Fig.~\ref{f: continuum}. $Q$ values less than 1 correspond to regions unstable against gravitational collapse.}
         \label{f: sim_Toomre}
   \end{figure}

In Fig.~\ref{f: sim_Toomre} we show the `true' $Q$ map obtained from the model simulations. As expected, we see low $Q$ values at the positions of the fragments since these regions have already collapsed to form protostars. There also exist low $Q$ values in a ring coincident with the edge of the disk. Since the mid-plane temperature is smoothly decreasing towards the edges of the disk (see Fig.~\ref{f: sim_temp}), one would expect a decrease in the $Q$ parameter in the outskirts of the disk, but the existence of low $Q$ values in a ring-like structure is due to enhanced disk surface density in a ring of material in this region (see Fig.~\ref{f: sim_sigma}). It is important to note that the Toomre analysis is meaningless outside of the disk region, therefore the transition from stable (high~$Q$) on envelope scales to unstable (low~$Q$) at larger radii in the outskirts of the maps has no physical meaning. This transition is best seen in the view closest to edge-on in Fig.~\ref{f: sim_Toomre} at a declination offset of about $\pm800$~au.

   \begin{figure*}
   \centering
   \includegraphics[width=\hsize]{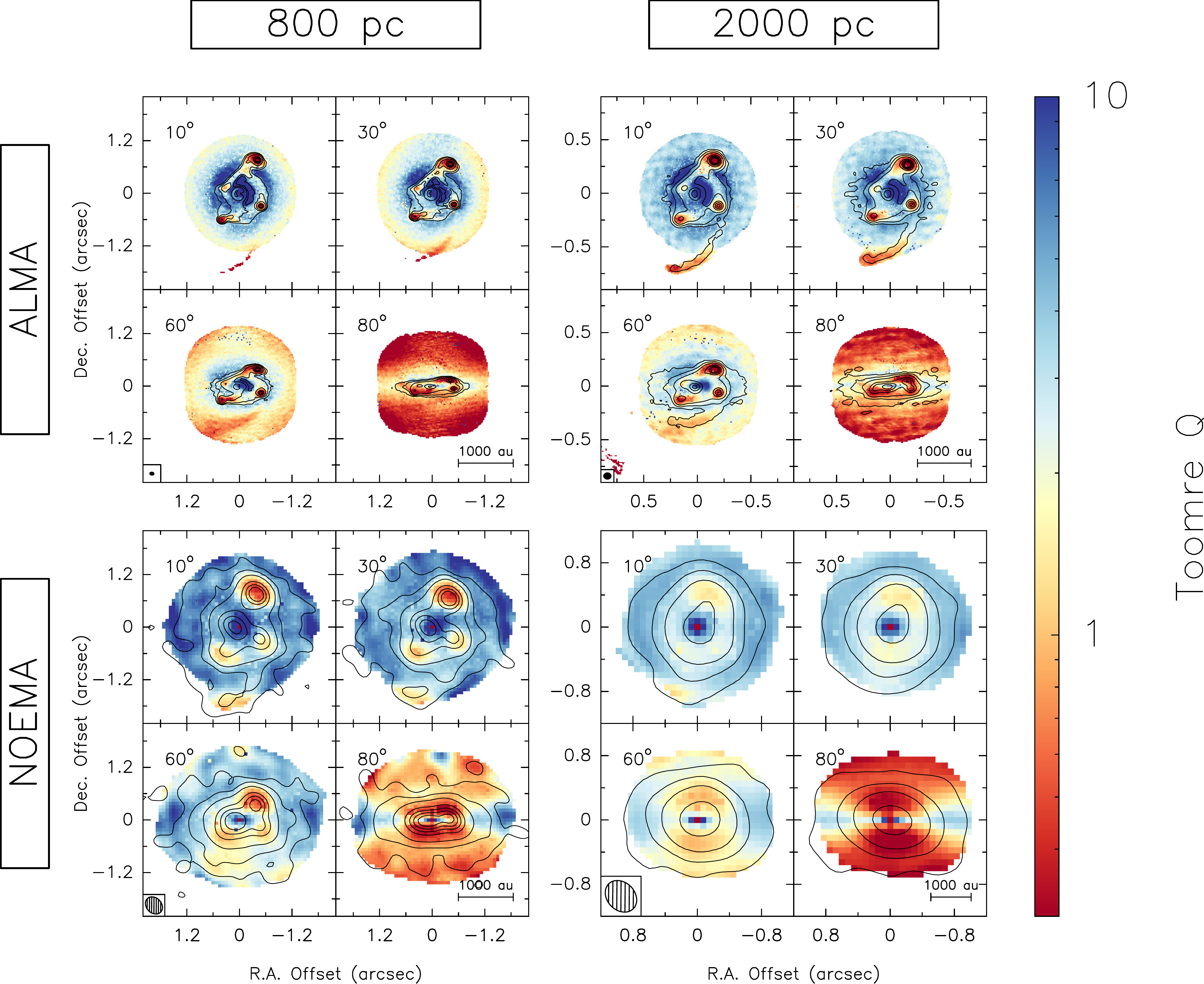}
      \caption{\tq\ map of synthetic ALMA observations \emph{(top row)}, and synthetic NOEMA observations \emph{(bottom row)} shifted to a distance of 800~pc (\emph{left column}), and 2000~pc (\emph{right column}). Each panel contains four sub-panels corresponding to the image inclined by 10$\degr$ \emph{(top left)}, 30$\degr$ \emph{(top right)}, 60$\degr$ \emph{(bottom left)}, and 80$\degr$ \emph{(bottom right)}. In each case, the disk is assumed to be in Keplerian rotation about a 10~\mo\ star at the center. The angular velocities have been corrected for the inclination (see text). The contours correspond to 1.37~mm continuum as shown in Fig.~\ref{f: continuum}. Synthesised beams are shown in the bottom left corners and scale bars are shown in the bottom right corners of each set of synthetic observations. $Q$ values less than 1 correspond to regions unstable against gravitational collapse.}
         \label{f: obs_Toomre_projected}
   \end{figure*}

In the following, we estimate the distribution of $Q$ for the synthetic observations and start the analysis by adopting the known protostellar mass at this snapshot and correcting the angular velocities for the known inclinations. The effects of not correcting the velocities properly and uncertainties in mass estimates are further discussed in Sections~\ref{sss: inclination_uncertainty} and \ref{sss: mass_uncertainty}, respectively. We assume that the LTE assumption is valid in such a high-density environment, and use the rotational temperature of \mc, modelled in Sect.~\ref{ss: temperature}, as representative of the kinetic temperature of the gas. To account for the self-gravity of the disk, we include the mass of the disk within a given radius, $M_\mathrm{disk}(r)$, to the mass of the protostar $M_\ast$ assumed to be 10~\mo. The angular velocity is further corrected to account for the inclination of the disk such that
\begin{equation} \label{e: omega}
\Omega(r_\mathrm{proj})=\sqrt{\frac{G\,(M_\ast+M_\mathrm{disk}(r_\mathrm{proj}))}{r_\mathrm{proj}^3}}.
\end{equation}
In particular, the radius at each pixel in a face-on disk is calculated as $r=\sqrt{\Delta x^2 +\Delta y^2}$ where $\Delta x$ is the distance from the protostar in right ascension and $\Delta y$ is the distance from the protostar in declination. Since we do not rotate the structure but only incline it about the $x$-axis, the projected radius is $r_\mathrm{proj}=\sqrt{\Delta x^2+\Delta y^2\,\mathrm{cos^2}(i)}$. 

The mass of the disk is calculated as discussed in Sect.~\ref{ss: core_mass} from the continuum emission maps, by summing up the mass contained within each projected radius using the aperture photometry tools of the \emph{photutils} package \citepads{2016ascl.soft09011B} within the \emph{astropy} package \citepads{2013A&A...558A..33A} in Python. Furthermore, the beam-averaged surface density of the disk is calculated via
\begin{equation}
\Sigma = \frac{S_\nu\,R}{B_\nu(T_D)\,\Omega_B\,\kappa_\nu},
\end{equation}
where $S_\nu$ is the peak intensity and $\Omega_B$ is the beam solid angle. 

In Fig.~\ref{f: obs_Toomre_projected}, we show the $Q$ maps for the synthetic observations. Again, as expected, we find low $Q$ values below the critical value of 1 at the positions of the fragments in the synthetic observations which resolve them. This is attributed to the enhanced surface density at the positions of the fragments which have already collapsed to form stars. We also find low $Q$ values ($\sim$1--2) in the arms connecting the fragments as well as the arc of material to the south being expelled from the system. The arms connecting the fragments can be locations where further disk fragmentation can take place. The extremely low $Q$ values above and below where the continuum contours are drawn in the view closest to edge-on (80\degr) are a result of low angular velocities at these positions, since the disk is only extending a small distance in declination. The gas contribution at these locations come from the rotating and infalling envelope and the Toomre analysis is actually not applicable for this envelope medium. 

The ring of low $Q$ values seen in the model simulations (see Fig.~\ref{f: sim_Toomre}) is not seen in the Toomre maps of the synthetic observations. Since the produced dust continuum images are essentially a convolution of the temperature with the density, the radial decrease of temperature in the outskirts (see Fig.~\ref{f: sim_temp}) where the ring exists in the density (see Fig.~\ref{f: sim_sigma}) results in this component not being prominent in the final continuum images (see Fig.~\ref{f: continuum}). Therefore, because we use the continuum maps to calculate the surface density of the disks in the Toomre equation, this component is also non-existent. Interestingly, we find high $Q$ values in a ring just outside of the lowest continuum contours, best seen in the synthetic ALMA observations inclined 10\degr\ and 30\degr. This is as a result of high gas temperatures in these regions, due to shock heating of gas falling from the rotating envelope onto the disk.

The synthetic NOEMA observations at 800~pc still show low $Q$ values at the positions of the fragments, but the low values expected to be seen in the arms connecting the fragments are absent because the temperature distribution is uniform in the inner regions (see bottom-left panel of Fig.~\ref{f: obs_temp}). Most interestingly, although the synthetic NOEMA observations at 2000~pc do not resolve the individual fragments, we find low $Q$ values $\sim$2 in the outskirts of the disk. This is as a result of the structure having a non-circular elongated shape in the direction of the brightest fragment in the north-west in both the continuum and temperature distributions. 

The critical value of $Q$ is often taken to be 1 in theoretical calculations, and lowers to $\sim$0.7 for an isothermal disk of finite thickness (\citeads{1965MNRAS.130..125G}; \citeads{2001ApJ...553..174G}). Considering the fact that our modelling of the level populations of the \mc\ $K$-ladder provides gas temperatures that may probe layers above the disk mid-plane, we assume that regions where $Q$ is less than $\sim$2 are unstable against gravitational collapse. The median $Q$ values are listed in Table~\ref{t: temp_mass_Toomre} and follow an opposite trend as the mass estimates. This is because the Toomre parameter is inversely proportional to the surface density of the disk which is calculated in a similar manner as the mass estimates but without the distance dependence. 

\begin{figure}
   \centering
   \includegraphics[width=\hsize]{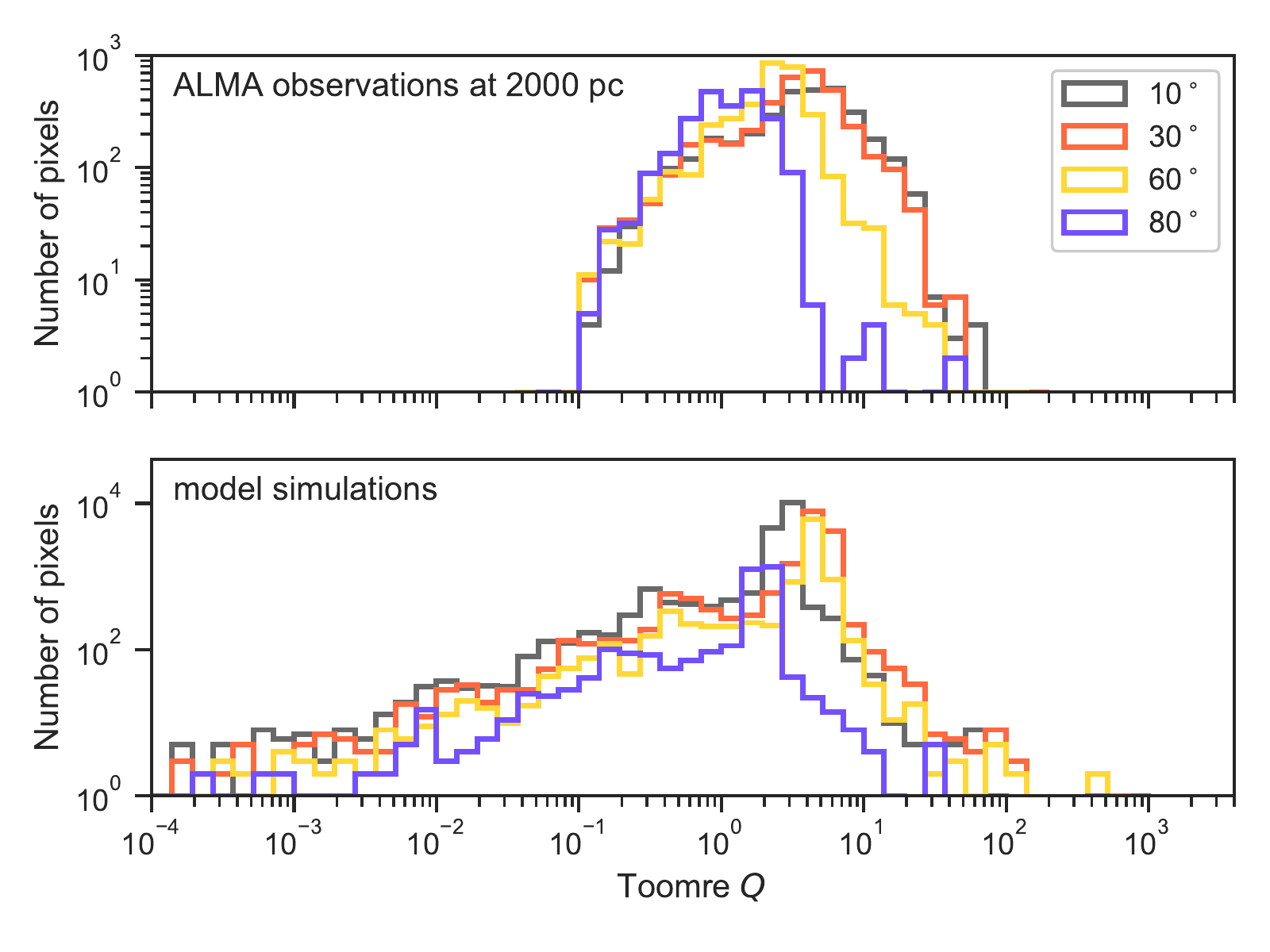}
      \caption{ \emph{Top:} Histogram of $Q$ values for synthetic ALMA observations at 2000~pc (see top-right panel of Fig.~\ref{f: obs_Toomre_projected}), including only pixels that lie within 6$\sigma$ continuum contours to show the $Q$ distribution mostly associated with the disk rather than the envelope. \emph{Bottom:} Histogram of $Q$ values for model simulations (see Fig.~\ref{f: sim_Toomre}), including only pixels within a radius of 1600~au.}
         \label{f: Toomre_histogram}
   \end{figure}

At first glance, one sees that on average, the $Q$ parameter gets lower with increasing inclination. This is naively expected as in the edge-on views one looks through a larger column. To investigate the effect of inclination further in the resolved observations, the histogram of $Q$ values for synthetic ALMA observations at 2000~pc is shown as an example in the top panels of Fig.~\ref{f: Toomre_histogram}. We attempt to remove as much contribution as possible from the envelope by only plotting the histogram of $Q$ values for the pixels that lie within 6$\sigma$ continuum contours which are the outer-most contours drawn on the maps in Fig.~\ref{f: obs_Toomre_projected}. The $Q$ distributions in these inner regions do not vary much for the 10\degr, 30\degr, and 60\degr\ views, but still a large portion of the contribution for the 80\degr\ inclined map comes from the envelope since the disk contribution is just a narrow region spanning a few pixels in declination. The angular velocities in the envelope for the case closest to edge-on are low and therefore result in low $Q$ values there. For comparison, in the bottom panel of Fig.~\ref{f: Toomre_histogram} we show the histogram of $Q$ values for the inclined models where only pixels within a radius of 1600~au are included in order to exclude the meaningless contribution from the larger-scale cloud. While the distribution is broader as there are more pixels with low~$Q$~values for each fragment, the findings remain the same. 

\subsubsection{Inclination of the angular velocity field} \label{sss: inclination_uncertainty}

In true observations, one seldom has information about the inclination of the source. Sometimes the outflow geometry can help, but high-mass stars form in highly clustered environments, and with the outflow angles widening with time \citepads{2005ASSL..324..105B}, there is a high chance that outflows emanating from different sources may overlap, making the deduction of an inclination angle for the disk difficult. To check how the lack of knowledge about the inclination of the disk would affect our analysis, we have produced an angular velocity map where at each pixel $r=\sqrt{\Delta x^2 +\Delta y^2}$, which essentially means we assume the angular velocity at each position is what it would be if the disk was edge-on in that direction. 

\begin{figure*}
   \centering
   \includegraphics[width=\hsize]{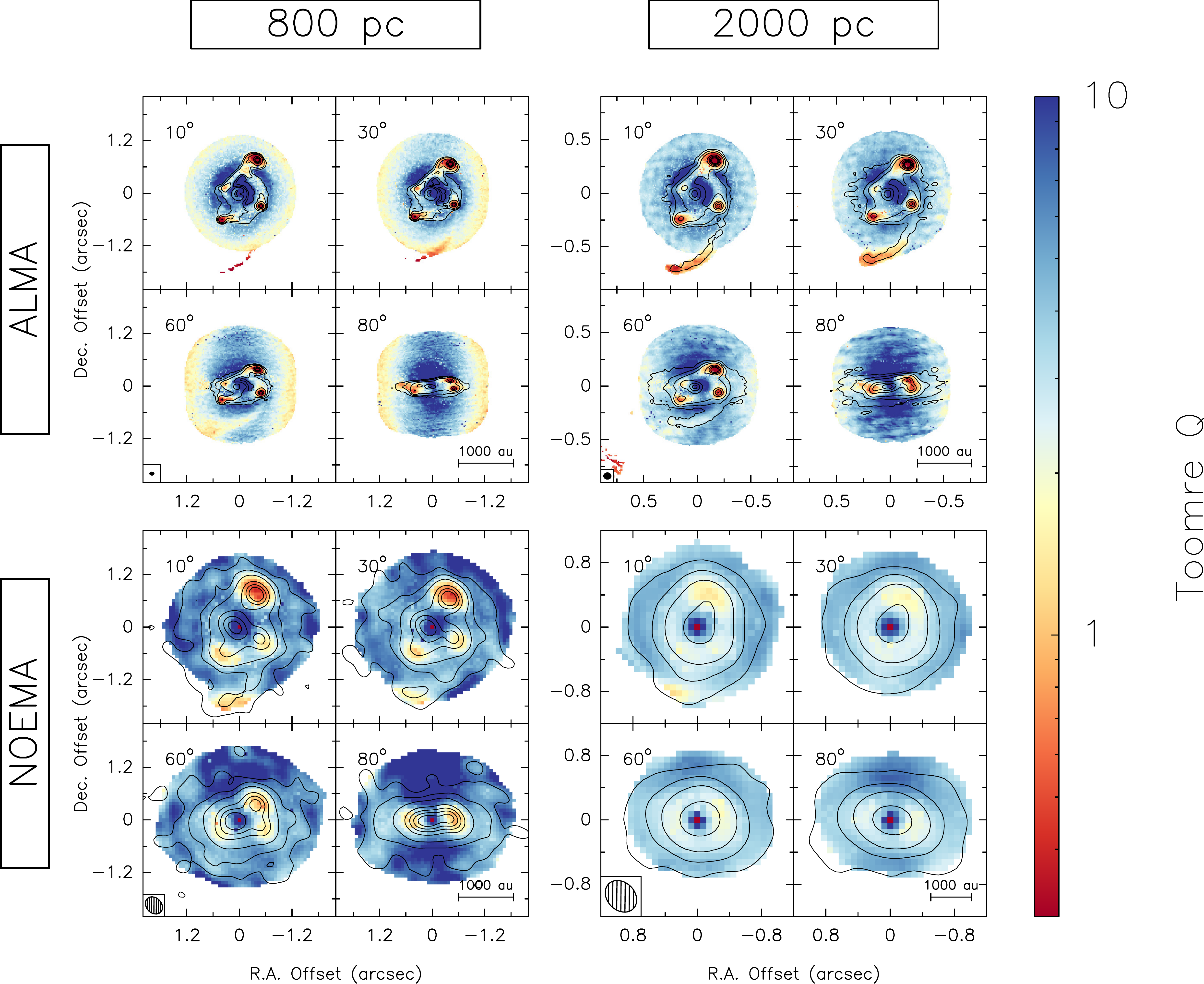}
      \caption{Same as Fig.~\ref{f: obs_Toomre_projected} but assuming the inclination of the object is not known, therefore, with the velocities not corrected to the actual rotation velocities.}
         \label{f: obs_Toomre}
   \end{figure*}

Figure~\ref{f: obs_Toomre} shows the $Q$ maps for the synthetic observations without the inclination correction. In the resolved ALMA observations at both distances and NOEMA observations at 800~pc, we still find low $Q$ values as expected at the positions of the fragments, but the higher inclined views show higher $Q$ values at the locations of the fragments than in the inclination-corrected cases, precisely because the angular velocities at a given point above and below the zero declination line are higher in the uncorrected maps. Interestingly, in the synthetic NOEMA observations at 2000~pc where the fragments are not resolved, we still see the low $Q$ values in the outskirts of the structure. Therefore, the conclusion from the corrected case still holds that we are able to predict fragmentation of the disk without actually resolving the fragments.

\subsubsection{Uncertainties in the mass of the central object} \label{sss: mass_uncertainty}
In our Toomre analysis, we have assumed the mass of the protostar at the center to be 10~\mo\ as we have access to this information. In Sect.~\ref{ss: pv_mass} we outlined how the mass of the central object is often estimated from the kinematics of the disk, and explained that with loss of resolution, the shape of the rotation curve of the disk changes from differential to a more rigid-body-like appearance. Therefore, fitting Keplerian-like curves to the PV plots of poorly-resolved structures would overestimate the mass of the central object. Since the mass of the protostar has a square root dependence in the expression of $Q$ (Eq.~\ref{e: Toomre}), overestimating the protostellar mass by a factor of 4 would yield $Q$ values that are on average higher by a factor of two, making the disks seem more stable. 

\section{Summary and Conclusions} \label{s: conclusions}
In this work, we have presented synthetic ALMA and NOEMA observations for a high resolution 3D radiation-hydrodynamic simulation of a high-mass protostellar disk that fragments into a highly dynamic system. At the snapshot of the simulation at 12~kyr, the system comprises of a central protostar and four companion fragments on scales $\leq$500~au, all accreting material from the infalling and rotating envelope which feeds the disk that is in Keplerian rotation. Smaller accretion disks are formed around each of the fragments, through which some of the large-scale disk and envelope material is accreted onto the fragments. The aim of the investigation has been to study the effect of inclination and spatial resolution in recovering disk structure and to study the stability of the disk in order to predict disk fragmentation and better understand recent and future high-resolution observations. The following is a summary of our findings:
\begin{itemize}
  \item Synthetic ALMA observation at 800~pc are able to detect all fragments at all inclinations in both continuum and line emission at 1.3~mm. At 2000~pc, the ALMA observations at 80\degr\ inclination cannot resolve the closest fragment to the protostar. The NOEMA observations at 800~pc can resolve all fragments at 10\degr\ and 30\degr\ inclinations but not all at higher inclinations. At 2000~pc, NOEMA observations only show a single structure, which looks slightly elongated towards the brightest fragment.
  \vspace{1mm}
  \item In the zeroth moment maps of \mc\ (12--11) rotational transitions, the peak of emission is at the position of the protostar. For the ALMA observations at both 800 and 2000~pc, we are able to detect the rotating envelope which is infalling onto a ring-like structure at the position of the centrifugal barrier. The effect is best seen in the 10\degr\ and 30\degr\ inclined views.  
  \vspace{1mm}
  \item In the first moment maps of \mc\ (12--11) rotational transitions, the disk contribution is best seen in the higher inclined views as expected. There exist small accretion disks around each fragment feeding them part of the disk material; this leads to the large-scale velocity field of the disk having a Yin-Yang shape. Envelope and disk contributions are heavily blended in the poorly resolved case of NOEMA observations at 2000~pc (linear resolutions $\sim$800~au) where the emission is also smeared over a larger distance, making the amplitude of the observed velocity gradient smaller. 
  \vspace{1mm}
  \item PV diagrams of  \mc\ (12--11) rotational transitions for a cut through the disk perpendicular to its rotation axis show the differential (Keplerian-like) rotation of material best in the close to edge-on view. At a radius of $\sim$250~au a discontinuity is seen, corresponding to the position of the centrifugal barrier because the infalling envelope has a higher infalling velocity than the radial velocity of the disk. As the entire structure becomes less resolved in the NOEMA observations, the high-velocity components of the disk become washed out and blended with the envelope component, making the PV diagrams resemble rigid-body-like rotation. 
  \vspace{1mm}
  \item We fit the emission of \mckr{4}{6} with \emph{XCLASS} to obtain rotational temperature maps for the disks. On average we find temperatures in the range of 100--300~K, decreasing as the system becomes more inclined. The envelope is roughly a factor of 2 cooler than the disk. In the poorly resolved NOEMA observations, the envelope and inner disk contributions become blended and the difference between the temperature of the inner and outer regions is not as apparent as in the resolved cases. 
  \vspace{1mm}
  \item Assuming gas and dust temperatures are coupled, mass estimates from dust for the synthetic observations account for roughly 20\% of the mass calculated from the model simulations for the inner 8400 by 8400~au and the rest is filtered out by the interferometers. 
  \vspace{1mm}
  \item We make use of the method presented in \citetads{2016MNRAS.459.1892S} to fit the PV plots with Keplerian models to obtain protostellar mass estimates. Since there exists a discontinuity in the 6$\sigma$ outer edge of the PV plots as a result of material falling from the envelope onto the disk, we fit the inner disk and outer envelope regions separately. In the case of well-resolved disks with Keplerian-like profiles, we find the fits to the inner regions the most accurate and the fits to the entire PV curves adequate for estimating enclosed masses. For poorly-resolved disks, the fits to the outer regions of the PV diagrams yield much more accurate mass estimates as opposed to fits to the entire region which overestimate the mass. The factor by which the mass is overestimated depends on how poorly the disk is resolved.
  \vspace{1mm}
  \item Studying the Toomre stability of the disks, we find low $Q$ values below the critical value for stability against gravitational collapse at the positions of the fragments and in the arms connecting the fragments for the resolved synthetic ALMA observations. For the synthetic NOEMA observations at 800~pc (linear resolutions $\sim$300~au), low $Q$ values are seen at the locations of the fragments but the unstable region in the disk is no longer seen since the temperature structure is uniformly warm for the inner disk. For the NOEMA observations at 2000~pc  (linear resolutions $\sim$800~au) where only one structure is resolved, we find low $Q$ values in the outskirts of the disk. This is an important result as it implies that despite the lack of ability in resolving any of the fragments, we are able to predict that the disk is unstable and fragmenting. 
  \vspace{1mm}
  \item Assuming that the inclination of the disk is not known, recalculating the $Q$ maps using Keplerian velocities of an edge-on disk to all radial directions for all inclinations provide slightly higher values for the $Q$ parameter but the conclusions from the previous point still hold. This is reassuring as disk inclinations are difficult to establish.
\end{itemize}

In this work, we showcased the potential and limitations to study disks in high-mass star formation with current (mm) interferometers. We benchmarked a method to study the stability of such disks, showing that even with poorly resolved observations it is possible to predict whether a disk is prone to fragmentation. We aim to apply our methods to the study of a large sample of high-mass disk candidates within the IRAM large program CORE in an upcoming paper. 

\begin{acknowledgements}
We thank the anonymous referee for providing detailed comments on the manuscript which helped with the clarity and quality of the work. AA and HB acknowledge support from the European Research Council under the European Community's Horizon 2020 framework program (2014-2020) via the ERC Consolidator Grant `From Cloud to Star Formation (CSF)' (project number 648505). RK acknowledges financial support via the Emmy Noether Research Group on Accretion Flows and Feedback in Realistic Models of Massive Star Formation funded by the German Research Foundation (DFG) under grant no. KU 2849/3-1 and KU 2849/3-2. 
\end{acknowledgements}


\begin{thebibliography}
\expandafter\ifx\csname natexlab\endcsname\relax\def\natexlab#1{#1}\fi

\bibitem[{{Ahmadi} {et~al.}(2018){Ahmadi}, {Beuther}, {Mottram}, {Bosco},
  {Linz}, {Henning}, {Winters}, {Kuiper}, {Pudritz}, {S{\'a}nchez-Monge},
  {Keto}, {Beltran}, {Bontemps}, {Cesaroni}, {Csengeri}, {Feng},
  {Galvan-Madrid}, {Johnston}, {Klaassen}, {Leurini}, {Longmore}, {Lumsden},
  {Maud}, {Menten}, {Moscadelli}, {Motte}, {Palau}, {Peters}, {Ragan},
  {Schilke}, {Urquhart}, {Wyrowski}, \& {Zinnecker}}]{2018A&A...618A..46A}
{Ahmadi}, A., {Beuther}, H., {Mottram}, J.~C., {et~al.} 2018, \aap, 618, A46

\bibitem[{{Alves} {et~al.}(2017){Alves}, {Girart}, {Caselli}, {Franco}, {Zhao},
  {Vlemmings}, {Evans}, \& {Ricci}}]{2017A&A...603L...3A}
{Alves}, F.~O., {Girart}, J.~M., {Caselli}, P., {et~al.} 2017, \aap, 603, L3

\bibitem[{{Astropy Collaboration} {et~al.}(2013){Astropy Collaboration},
  {Robitaille}, {Tollerud}, {Greenfield}, {Droettboom}, {Bray}, {Aldcroft},
  {Davis}, {Ginsburg}, {Price-Whelan}, {Kerzendorf}, {Conley}, {Crighton},
  {Barbary}, {Muna}, {Ferguson}, {Grollier}, {Parikh}, {Nair}, {Unther},
  {Deil}, {Woillez}, {Conseil}, {Kramer}, {Turner}, {Singer}, {Fox}, {Weaver},
  {Zabalza}, {Edwards}, {Azalee Bostroem}, {Burke}, {Casey}, {Crawford},
  {Dencheva}, {Ely}, {Jenness}, {Labrie}, {Lim}, {Pierfederici}, {Pontzen},
  {Ptak}, {Refsdal}, {Servillat}, \& {Streicher}}]{2013A&A...558A..33A}
{Astropy Collaboration}, {Robitaille}, T.~P., {Tollerud}, E.~J., {et~al.} 2013,
  \aap, 558, A33

\bibitem[{{Banerjee} \& {Pudritz}(2008)}]{2008ASPC..387..216B}
{Banerjee}, R. \& {Pudritz}, R.~E. 2008, in Astronomical Society of the Pacific
  Conference Series, Vol. 387, Massive Star Formation: Observations Confront
  Theory, ed. H.~{Beuther}, H.~{Linz}, \& T.~{Henning}, 216

\bibitem[{{Beltr{\'a}n} \& {de Wit}(2016)}]{2016A&ARv..24....6B}
{Beltr{\'a}n}, M.~T. \& {de Wit}, W.~J. 2016, Astronomy and Astrophysics
  Review, 24, 6

\bibitem[{{Beuther} {et~al.}(2018){Beuther}, {Mottram}, {Ahmadi}, {Bosco},
  {Linz}, {Henning}, {Klaassen}, {Winters}, {Maud}, {Kuiper}, {Semenov},
  {Gieser}, {Peters}, {Urquhart}, {Pudritz}, {Ragan}, {Feng}, {Keto},
  {Leurini}, {Cesaroni}, {Beltran}, {Palau}, {S{\'a}nchez-Monge},
  {Galvan-Madrid}, {Zhang}, {Schilke}, {Wyrowski}, {Johnston}, {Longmore},
  {Lumsden}, {Hoare}, {Menten}, \& {Csengeri}}]{2018A&A...617A.100B}
{Beuther}, H., {Mottram}, J.~C., {Ahmadi}, A., {et~al.} 2018, \aap, 617, A100

\bibitem[{{Beuther} {et~al.}(2002){Beuther}, {Schilke}, {Gueth}, {McCaughrean},
  {Andersen}, {Sridharan}, \& {Menten}}]{2002A&A...387..931B}
{Beuther}, H., {Schilke}, P., {Gueth}, F., {et~al.} 2002, \aap, 387, 931

\bibitem[{{Beuther} \& {Shepherd}(2005)}]{2005ASSL..324..105B}
{Beuther}, H. \& {Shepherd}, D. 2005, in Astrophysics and Space Science
  Library, Vol. 324, Astrophysics and Space Science Library, ed. M.~S.~N.
  {Kumar}, M.~{Tafalla}, \& P.~{Caselli}, 105

\bibitem[{Bhandare {et~al.}(2018)Bhandare, Kuiper, Henning, Fendt, Marleau, \&
  K{\"o}lligan}]{2018A&A...618A..95B}
Bhandare, A., Kuiper, R., Henning, T., {et~al.} 2018, arXiv, A95

\bibitem[{{Bosco} {et~al.}(2019){Bosco}, {Beuther}, {Ahmadi}, {Mottram},
  {Kuiper}, {Linz}, {Maud}, {Winters}, {Henning}, {Feng}, {Peters}, {Semenov},
  {Klaassen}, {Schilke}, {Urquhart}, {Beltr{\'a}n}, {Lumsden}, {Leurini},
  {Moscadelli}, {Cesaroni}, {S{\'a}nchez-Monge}, {Palau}, {Pudritz},
  {Wyrowski}, \& {Longmore}}]{2019A&A...629A..10B}
{Bosco}, F., {Beuther}, H., {Ahmadi}, A., {et~al.} 2019, \aap, 629, A10

\bibitem[{{Bradley} {et~al.}(2016){Bradley}, {Sipocz}, {Robitaille},
  {Tollerud}, {Deil}, {Vin{\'\i}cius}, {Barbary}, {G{\"u}nther}, {Bostroem},
  {Droettboom}, {Bray}, {Bratholm}, {Pickering}, {Craig}, {Pascual}, {Greco},
  {Donath}, {Kerzendorf}, {Littlefair}, {Barentsen}, {D'Eugenio}, \&
  {Weaver}}]{2016ascl.soft09011B}
{Bradley}, L., {Sipocz}, B., {Robitaille}, T., {et~al.} 2016, {Photutils:
  Photometry tools}

\bibitem[{{Cesaroni} {et~al.}(2017){Cesaroni}, {S{\'a}nchez-Monge},
  {Beltr{\'a}n}, {Johnston}, {Maud}, {Moscadelli}, {Mottram}, {Ahmadi},
  {Allen}, {Beuther}, {Csengeri}, {Etoka}, {Fuller}, {Galli},
  {Galv{\'a}n-Madrid}, {Goddi}, {Henning}, {Hoare}, {Klaassen}, {Kuiper},
  {Kumar}, {Lumsden}, {Peters}, {Rivilla}, {Schilke}, {Testi}, {van der Tak},
  {Vig}, {Walmsley}, \& {Zinnecker}}]{2017A&A...602A..59C}
{Cesaroni}, R., {S{\'a}nchez-Monge}, {\'A}., {Beltr{\'a}n}, M.~T., {et~al.}
  2017, \aap, 602, A59

\bibitem[{{Chabrier}(2003)}]{2003PASP..115..763C}
{Chabrier}, G. 2003, Publications of the Astronomical Society of the Pacific,
  115, 763

\bibitem[{{Chen} {et~al.}(2016){Chen}, {Keto}, {Zhang}, {Sridharan}, {Liu}, \&
  {Su}}]{2016ApJ...823..125C}
{Chen}, H.-R.~V., {Keto}, E., {Zhang}, Q., {et~al.} 2016, \apj, 823, 125

\bibitem[{{Chini} {et~al.}(2012){Chini}, {Hoffmeister}, {Nasseri}, {Stahl}, \&
  {Zinnecker}}]{2012MNRAS.424.1925C}
{Chini}, R., {Hoffmeister}, V.~H., {Nasseri}, A., {Stahl}, O., \& {Zinnecker},
  H. 2012, \mnras, 424, 1925

\bibitem[{{Clark}(1980)}]{1980A&A....89..377C}
{Clark}, B.~G. 1980, \aap, 89, 377

\bibitem[{{Collings} {et~al.}(2004){Collings}, {Anderson}, {Chen}, {Dever},
  {Viti}, {Williams}, \& {McCoustra}}]{2004MNRAS.354.1133C}
{Collings}, M.~P., {Anderson}, M.~A., {Chen}, R., {et~al.} 2004, \mnras, 354,
  1133

\bibitem[{Commer{\c c}on {et~al.}(2011)Commer{\c c}on, Teyssier, Audit,
  Hennebelle, \& Chabrier}]{2011A&A...529A..35C}
Commer{\c c}on, B., Teyssier, R., Audit, E., Hennebelle, P., \& Chabrier, G.
  2011, A{\&}A, 529, A35

\bibitem[{Courant {et~al.}(1967)Courant, Friedrichs, \&
  Lewy}]{1967IBMJ...11..215C}
Courant, R., Friedrichs, K., \& Lewy, H. 1967, IBM Journal of Research and
  Development, 11, 215

\bibitem[{{Csengeri} {et~al.}(2018){Csengeri}, {Bontemps}, {Wyrowski},
  {Belloche}, {Menten}, {Leurini}, {Beuther}, {Bronfman}, {Commer{\c{c}}on},
  {Chapillon}, {Longmore}, {Palau}, {Tan}, \& {Urquhart}}]{2018A&A...617A..89C}
{Csengeri}, T., {Bontemps}, S., {Wyrowski}, F., {et~al.} 2018, \aap, 617, A89

\bibitem[{{Draine}(2011)}]{2011piim.book.....D}
{Draine}, B.~T. 2011, {Physics of the Interstellar and Intergalactic Medium}
  (Princeton University Press)

\bibitem[{Dullemond(2012)}]{2012ascl.soft02015D}
Dullemond, C.~P. 2012, ASCL, ascl:1202.015

\bibitem[{{Fallscheer} {et~al.}(2009){Fallscheer}, {Beuther}, {Zhang}, {Keto},
  \& {Sridharan}}]{2009A&A...504..127F}
{Fallscheer}, C., {Beuther}, H., {Zhang}, Q., {Keto}, E., \& {Sridharan}, T.~K.
  2009, \aap, 504, 127

\bibitem[{{Frank} {et~al.}(2014){Frank}, {Ray}, {Cabrit}, {Hartigan}, {Arce},
  {Bacciotti}, {Bally}, {Benisty}, {Eisl{\"o}ffel}, {G{\"u}del}, {Lebedev},
  {Nisini}, \& {Raga}}]{2014prpl.conf..451F}
{Frank}, A., {Ray}, T.~P., {Cabrit}, S., {et~al.} 2014, in Protostars and
  Planets VI, ed. H.~{Beuther}, R.~{Klessen}, C.~{Dullemond}, \& T.~{Henning}
  (Univ. of Arizona Press, Tucson), 451--474

\bibitem[{{Gammie}(2001)}]{2001ApJ...553..174G}
{Gammie}, C.~F. 2001, \apj, 553, 174

\bibitem[{{Gerner} {et~al.}(2014){Gerner}, {Beuther}, {Semenov}, {Linz},
  {Vasyunina}, {Bihr}, {Shirley}, \& {Henning}}]{2014A&A...563A..97G}
{Gerner}, T., {Beuther}, H., {Semenov}, D., {et~al.} 2014, \aap, 563, A97

\bibitem[{{Goldreich} \& {Lynden-Bell}(1965)}]{1965MNRAS.130..125G}
{Goldreich}, P. \& {Lynden-Bell}, D. 1965, \mnras, 130, 125

\bibitem[{{Harries} {et~al.}(2017){Harries}, {Douglas}, \&
  {Ali}}]{2017MNRAS.471.4111H}
{Harries}, T.~J., {Douglas}, T.~A., \& {Ali}, A. 2017, \mnras, 471, 4111

\bibitem[{{Haworth} {et~al.}(2018){Haworth}, {Glover}, {Koepferl}, {Bisbas}, \&
  {Dale}}]{2018NewAR..82....1H}
{Haworth}, T.~J., {Glover}, S. C.~O., {Koepferl}, C.~M., {Bisbas}, T.~G., \&
  {Dale}, J.~E. 2018, New Astronomy Reviews, 82, 1

\bibitem[{{Hildebrand}(1983)}]{1983QJRAS..24..267H}
{Hildebrand}, R.~H. 1983, \qjras, 24, 267

\bibitem[{{H{\"o}gbom}(1974)}]{1974A&AS...15..417H}
{H{\"o}gbom}, J.~A. 1974, Astronomy and Astrophysics Supplement Series, 15, 417

\bibitem[{Hosokawa \& Omukai(2009)}]{2009ApJ...691..823H}
Hosokawa, T. \& Omukai, K. 2009, ApJ, 691, 823

\bibitem[{{Ilee} {et~al.}(2018){Ilee}, {Cyganowski}, {Brogan}, {Hunter},
  {Forgan}, {Haworth}, {Clarke}, \& {Harries}}]{2018ApJ...869L..24I}
{Ilee}, J.~D., {Cyganowski}, C.~J., {Brogan}, C.~L., {et~al.} 2018, \apj, 869,
  L24

\bibitem[{{Ilee} {et~al.}(2016){Ilee}, {Cyganowski}, {Nazari}, {Hunter},
  {Brogan}, {Forgan}, \& {Zhang}}]{2016MNRAS.462.4386I}
{Ilee}, J.~D., {Cyganowski}, C.~J., {Nazari}, P., {et~al.} 2016, \mnras, 462,
  4386

\bibitem[{{Jankovic} {et~al.}(2019){Jankovic}, {Haworth}, {Ilee}, {Forgan},
  {Cyganowski}, {Walsh}, {Brogan}, {Hunter}, \&
  {Mohanty}}]{2019MNRAS.482.4673J}
{Jankovic}, M.~R., {Haworth}, T.~J., {Ilee}, J.~D., {et~al.} 2019, \mnras, 482,
  4673

\bibitem[{{Johnston} {et~al.}(2015){Johnston}, {Robitaille}, {Beuther}, {Linz},
  {Boley}, {Kuiper}, {Keto}, {Hoare}, \& {van Boekel}}]{2015ApJ...813L..19J}
{Johnston}, K.~G., {Robitaille}, T.~P., {Beuther}, H., {et~al.} 2015, \apjl,
  813, L19

\bibitem[{{Kahn}(1974)}]{1974A&A....37..149K}
{Kahn}, F.~D. 1974, \aap, 37, 149

\bibitem[{{Klassen} {et~al.}(2016){Klassen}, {Pudritz}, {Kuiper}, {Peters}, \&
  {Banerjee}}]{2016ApJ...823...28K}
{Klassen}, M., {Pudritz}, R.~E., {Kuiper}, R., {Peters}, T., \& {Banerjee}, R.
  2016, \apj, 823, 28

\bibitem[{{K{\"o}lligan} \& {Kuiper}(2018)}]{2018A&A...620A.182K}
{K{\"o}lligan}, A. \& {Kuiper}, R. 2018, \aap, 620, A182

\bibitem[{{Kratter} \& {Matzner}(2006)}]{2006MNRAS.373.1563K}
{Kratter}, K.~M. \& {Matzner}, C.~D. 2006, \mnras, 373, 1563

\bibitem[{{Kratter} {et~al.}(2010){Kratter}, {Matzner}, {Krumholz}, \&
  {Klein}}]{2010ApJ...708.1585K}
{Kratter}, K.~M., {Matzner}, C.~D., {Krumholz}, M.~R., \& {Klein}, R.~I. 2010,
  \apj, 708, 1585

\bibitem[{{Kroupa}(2001)}]{2001MNRAS.322..231K}
{Kroupa}, P. 2001, \mnras, 322, 231

\bibitem[{{Krumholz} {et~al.}(2007){Krumholz}, {Klein}, \&
  {McKee}}]{2007ApJ...665..478K}
{Krumholz}, M.~R., {Klein}, R.~I., \& {McKee}, C.~F. 2007, \apj, 665, 478

\bibitem[{{Krumholz} {et~al.}(2009){Krumholz}, {Klein}, {McKee}, {Offner}, \&
  {Cunningham}}]{2009Sci...323..754K}
{Krumholz}, M.~R., {Klein}, R.~I., {McKee}, C.~F., {Offner}, S.~S.~R., \&
  {Cunningham}, A.~J. 2009, Science, 323, 754

\bibitem[{{Kuiper} \& {Hosokawa}(2018)}]{2018A&A...616A.101K}
{Kuiper}, R. \& {Hosokawa}, T. 2018, \aap, 616, A101

\bibitem[{{Kuiper} {et~al.}(2010){Kuiper}, {Klahr}, {Beuther}, \&
  {Henning}}]{2010ApJ...722.1556K}
{Kuiper}, R., {Klahr}, H., {Beuther}, H., \& {Henning}, T. 2010, \apj, 722,
  1556

\bibitem[{{Kuiper} {et~al.}(2011){Kuiper}, {Klahr}, {Beuther}, \&
  {Henning}}]{2011ApJ...732...20K}
{Kuiper}, R., {Klahr}, H., {Beuther}, H., \& {Henning}, T. 2011, \apj, 732, 20

\bibitem[{Kuiper {et~al.}(2010)Kuiper, Klahr, Dullemond, Kley, \&
  Henning}]{2010A&A...511A..81K}
Kuiper, R., Klahr, H., Dullemond, C., Kley, W., \& Henning, T. 2010, A{\&}A,
  511, A81

\bibitem[{Kuiper \& Klessen(2013)}]{2013A&A...555A...7K}
Kuiper, R. \& Klessen, R.~S. 2013, A{\&}A, 555, A7

\bibitem[{{Kuiper} {et~al.}(2016){Kuiper}, {Turner}, \&
  {Yorke}}]{2016ApJ...832...40K}
{Kuiper}, R., {Turner}, N.~J., \& {Yorke}, H.~W. 2016, \apj, 832, 40

\bibitem[{{Kuiper} \& {Yorke}(2013)}]{2013ApJ...772...61K}
{Kuiper}, R. \& {Yorke}, H.~W. 2013, \apj, 772, 61

\bibitem[{Kuiper {et~al.}(2015)Kuiper, Yorke, \& Turner}]{2015ApJ...800...86K}
Kuiper, R., Yorke, H.~W., \& Turner, N.~J. 2015, ApJ, 800, 86

\bibitem[{Laor \& Draine(1993)}]{1993ApJ...402..441L}
Laor, A. \& Draine, B.~T. 1993, ApJ, 402, 441

\bibitem[{{Larson} \& {Starrfield}(1971)}]{1971A&A....13..190L}
{Larson}, R.~B. \& {Starrfield}, S. 1971, \aap, 13, 190

\bibitem[{{Lee} {et~al.}(2017){Lee}, {Ho}, {Li}, {Hirano}, {Zhang}, \&
  {Shang}}]{2017NatAs...1E.152L}
{Lee}, C.-F., {Ho}, P. T.~P., {Li}, Z.-Y., {et~al.} 2017, Nature Astronomy, 1,
  0152

\bibitem[{{Leurini} {et~al.}(2011){Leurini}, {Codella}, {Zapata},
  {Beltr{\'a}n}, {Schilke}, \& {Cesaroni}}]{2011A&A...530A..12L}
{Leurini}, S., {Codella}, C., {Zapata}, L., {et~al.} 2011, \aap, 530, A12

\bibitem[{{Maud} {et~al.}(2019){Maud}, {Cesaroni}, {Kumar}, {Rivilla},
  {Ginsburg}, {Klaassen}, {Harsono}, {S{\'a}nchez-Monge}, {Ahmadi}, {Allen},
  {Beltr{\'a}n}, {Beuther}, {Galv{\'a}n-Madrid}, {Goddi}, {Hoare},
  {Hogerheijde}, {Johnston}, {Kuiper}, {Moscadelli}, {Peters}, {Testi}, {van
  der Tak}, \& {de Wit}}]{2019A&A...627L...6M}
{Maud}, L.~T., {Cesaroni}, R., {Kumar}, M.~S.~N., {et~al.} 2019, \aap, 627, L6

\bibitem[{{Maud} {et~al.}(2018){Maud}, {Cesaroni}, {Kumar}, {van der Tak},
  {Allen}, {Hoare}, {Klaassen}, {Harsono}, {Hogerheijde}, {S{\'a}nchez-Monge},
  {Schilke}, {Ahmadi}, {Beltr{\'a}n}, {Beuther}, {Csengeri}, {Etoka}, {Fuller},
  {Galv{\'a}n-Madrid}, {Goddi}, {Henning}, {Johnston}, {Kuiper}, {Lumsden},
  {Moscadelli}, {Mottram}, {Peters}, {Rivilla}, {Testi}, {Vig}, {de Wit}, \&
  {Zinnecker}}]{2018A&A...620A..31M}
{Maud}, L.~T., {Cesaroni}, R., {Kumar}, M.~S.~N., {et~al.} 2018, \aap, 620, A31

\bibitem[{{Maud} {et~al.}(2015){Maud}, {Moore}, {Lumsden}, {Mottram},
  {Urquhart}, \& {Hoare}}]{2015MNRAS.453..645M}
{Maud}, L.~T., {Moore}, T.~J.~T., {Lumsden}, S.~L., {et~al.} 2015, \mnras, 453,
  645

\bibitem[{{Meyer} {et~al.}(2018){Meyer}, {Kuiper}, {Kley}, {Johnston}, \&
  {Vorobyov}}]{2018MNRAS.473.3615M}
{Meyer}, D.~M.~A., {Kuiper}, R., {Kley}, W., {Johnston}, K.~G., \& {Vorobyov},
  E. 2018, \mnras, 473, 3615

\bibitem[{{Meyer} {et~al.}(2017){Meyer}, {Vorobyov}, {Kuiper}, \&
  {Kley}}]{2017MNRAS.464L..90M}
{Meyer}, D.~M.-A., {Vorobyov}, E.~I., {Kuiper}, R., \& {Kley}, W. 2017, \mnras,
  464, L90

\bibitem[{Mignone {et~al.}(2007)Mignone, Bodo, Massaglia, Matsakos, Tesileanu,
  Zanni, \& Ferrari}]{2007ApJS..170..228M}
Mignone, A., Bodo, G., Massaglia, S., {et~al.} 2007, ApJS, 170, 228

\bibitem[{Mignone {et~al.}(2012)Mignone, Zanni, Tzeferacos, Van~Straalen,
  Colella, \& Bodo}]{2012ApJS..198....7M}
Mignone, A., Zanni, C., Tzeferacos, P., {et~al.} 2012, ApJS, 198, 7

\bibitem[{{M{\"o}ller} {et~al.}(2017){M{\"o}ller}, {Endres}, \&
  {Schilke}}]{2017A&A...598A...7M}
{M{\"o}ller}, T., {Endres}, C., \& {Schilke}, P. 2017, \aap, 598, A7

\bibitem[{{Motte} {et~al.}(2018){Motte}, {Bontemps}, \&
  {Louvet}}]{2018ARA&A..56...41M}
{Motte}, F., {Bontemps}, S., \& {Louvet}, F. 2018, Annual Review of Astronomy
  and Astrophysics, 56, 41

\bibitem[{{Nakano}(1989)}]{1989ApJ...345..464N}
{Nakano}, T. 1989, \apj, 345, 464

\bibitem[{{Ossenkopf} \& {Henning}(1994)}]{1994A&A...291..943O}
{Ossenkopf}, V. \& {Henning}, T. 1994, \aap, 291, 943

\bibitem[{{Oya} {et~al.}(2018){Oya}, {Moriwaki}, {Onishi}, {Sakai},
  {L{\'o}pez─Sepulcre}, {Favre}, {Watanabe}, {Ceccarelli}, {Lefloch}, \&
  {Yamamoto}}]{2018ApJ...854...96O}
{Oya}, Y., {Moriwaki}, K., {Onishi}, S., {et~al.} 2018, \apj, 854, 96

\bibitem[{{Oya} {et~al.}(2016){Oya}, {Sakai}, {L{\'o}pez-Sepulcre}, {Watanabe},
  {Ceccarelli}, {Lefloch}, {Favre}, \& {Yamamoto}}]{2016ApJ...824...88O}
{Oya}, Y., {Sakai}, N., {L{\'o}pez-Sepulcre}, A., {et~al.} 2016, \apj, 824, 88

\bibitem[{{Palla} \& {Stahler}(1993)}]{1993ApJ...418..414P}
{Palla}, F. \& {Stahler}, S.~W. 1993, \apj, 418, 414

\bibitem[{{Pudritz} {et~al.}(2007){Pudritz}, {Ouyed}, {Fendt}, \&
  {Brandenburg}}]{2007prpl.conf..277P}
{Pudritz}, R.~E., {Ouyed}, R., {Fendt}, C., \& {Brandenburg}, A. 2007, in
  Protostars and Planets V, ed. B.~{Reipurth}, D.~{Jewitt}, \& K.~{Keil} (Univ.
  of Arizona Press, Tucson), 277--294

\bibitem[{{Rosen} {et~al.}(2016){Rosen}, {Krumholz}, {McKee}, \&
  {Klein}}]{2016MNRAS.463.2553R}
{Rosen}, A.~L., {Krumholz}, M.~R., {McKee}, C.~F., \& {Klein}, R.~I. 2016,
  \mnras, 463, 2553

\bibitem[{{Safronov}(1960)}]{1960AnAp...23..979S}
{Safronov}, V.~S. 1960, Annales d'Astrophysique, 23, 979

\bibitem[{{Sakai} {et~al.}(2014){Sakai}, {Sakai}, {Hirota}, {Watanabe},
  {Ceccarelli}, {Kahane}, {Bottinelli}, {Caux}, {Demyk}, {Vastel}, {Coutens},
  {Taquet}, {Ohashi}, {Takakuwa}, {Yen}, {Aikawa}, \&
  {Yamamoto}}]{2014Natur.507...78S}
{Sakai}, N., {Sakai}, T., {Hirota}, T., {et~al.} 2014, \nat, 507, 78

\bibitem[{{Sanna} {et~al.}(2019){Sanna}, {K{\"o}lligan}, {Moscadelli},
  {Kuiper}, {Cesaroni}, {Pillai}, {Menten}, {Zhang}, {Caratti o Garatti},
  {Goddi}, {Leurini}, \& {Carrasco-Gonz{\'a}lez}}]{2019A&A...623A..77S}
{Sanna}, A., {K{\"o}lligan}, A., {Moscadelli}, L., {et~al.} 2019, \aap, 623,
  A77

\bibitem[{{Seifried} {et~al.}(2016){Seifried}, {S{\'a}nchez-Monge}, {Walch}, \&
  {Banerjee}}]{2016MNRAS.459.1892S}
{Seifried}, D., {S{\'a}nchez-Monge}, {\'A}., {Walch}, S., \& {Banerjee}, R.
  2016, \mnras, 459, 1892

\bibitem[{{Toomre}(1964)}]{1964ApJ...139.1217T}
{Toomre}, A. 1964, \apj, 139, 1217

\bibitem[{{Wolfire} \& {Cassinelli}(1987)}]{1987ApJ...319..850W}
{Wolfire}, M.~G. \& {Cassinelli}, J.~P. 1987, \apj, 319, 850

\bibitem[{{Yorke} \& {Sonnhalter}(2002)}]{2002ApJ...569..846Y}
{Yorke}, H.~W. \& {Sonnhalter}, C. 2002, \apj, 569, 846

\bibitem[{{Zapata} {et~al.}(2019){Zapata}, {Garay}, {Palau},
  {Rodr{\'{\i}}guez}, {Fern{\'a}ndez-L{\'o}pez}, {Estalella}, \&
  {Guzm{\'a}n}}]{2019ApJ...872..176Z}
{Zapata}, L.~A., {Garay}, G., {Palau}, A., {et~al.} 2019, \apj, 872, 176

\bibitem[{{Zapata} {et~al.}(2015){Zapata}, {Palau}, {Galv{\'a}n-Madrid},
  {Rodr{\'{\i}}guez}, {Garay}, {Moran}, \&
  {Franco-Hern{\'a}ndez}}]{2015MNRAS.447.1826Z}
{Zapata}, L.~A., {Palau}, A., {Galv{\'a}n-Madrid}, R., {et~al.} 2015, \mnras,
  447, 1826

\end{thebibliography}
\end{document}